\documentclass{aa}

\usepackage{graphicx,xcolor}
\usepackage{txfonts}
\usepackage{arydshln}
\usepackage[normalem]{ulem}
\usepackage[dvipsnames]{xcolor}
\usepackage[table]{xcolor}
\usepackage{multirow}
\def\orb{\mathrm{orb}}
\def\sys{\mathrm{sys}}
\def\env{\mathrm{env}}
\def\drag{\mathrm{drag}}
\def\disrupt{\mathrm{disrupt}}
\def\Roche{\mathrm{Roche}}

\def\Msun{\,M_\odot}
\def\Mj{M_J}
\def\Rj{R_J}
\def\wind{\mathrm{wind}}
\def\engl{\mathrm{engl}}
\def\bind{\mathrm{bind}}
\def\eqp{\mathrm{eq},p}
\def\eqs{\mathrm{eq},\star}

\def\eject{\mathrm{eject}}

\begin{document} 
    %\selectlanguage{UKenglish}
   \title{On the formation of subdwarf B stars via engulfment \\ of substellar companions}
   \subtitle{I. Conditions for and occurrence of envelope ejection events}
   
   \author{A. Thuillier \inst{1, 2, *}
          \and
          L. Siess \inst{2}
          \and
          V. Van Grootel \inst{1}
          %\and Fran mmmm
          }

    \institute{
    Space sciences, Technologies and Astrophysics Research (STAR) Institute, Université de Liège, 19C Allée du 6 Août, 4000 Liège, Belgium\\
    \and
    Institut d’Astronomie et d’Astrophysique, Université Libre de Bruxelles (ULB), CP 226 1050 Bruxelles, Belgium\\
    * \email{antoine.thuillier@uliege.be}}

   \date{Received ******; accepted ******}
 
  \abstract
  % context heading (optional)
   {The canonical scenario for the formation of subdwarf B (sdB) stars involves the ejection of the progenitor's envelope near the tip of the Red Giant Branch (RGB), concurrent with the onset of core He-burning. While binary interactions are known to dominate sdB formation, the origin of apparently single sdB stars remains uncertain.}
  % aims heading (mandatory)
   {We investigate the conditions under which an sdB progenitor can eject its envelope through the engulfment of a substellar companion (planet or brown dwarf), using an energy balance approach.}
  % methods heading (mandatory)
   {We simulate the orbital evolution of substellar companions (1–80 $\Mj$) during the subgiant and RGB phases of their host stars (1.2-2.0 $\Msun$) to determine the onset of engulfment and calculate the energy released during inspiral within the stellar envelope. By comparing this energy with the envelope binding energy and exploring different efficiencies for its deposition into the envelope, we identify the conditions required for envelope ejection. We then combine these results with the observed population of substellar companions to estimate the occurrence of envelope-ejection events.}
  % results heading (mandatory)
   {The engulfment of substellar objects can lead to envelope ejection within a limited region of the mass-semi-major axis parameter space, whose extent depends on the host star's properties  and energy transfer efficiency $\alpha$. The minimum companion mass required for ejection increases with decreasing efficiency and increasing stellar mass. As a result, the frequency of envelope ejection is strongly influenced by the distribution of substellar companions, particularly by the paucity of objects in the brown-dwarf desert.}
  % conclusions heading (optional), leave it empty if necessary 
   {The engulfment of low-mass brown dwarfs and massive planets during the late RGB can provide sufficient energy to eject the stellar envelope, which could ultimately lead to the formation of sdB stars. Our results define the region of parameter space where this mechanism is energetically possible and provide a guide for future multidimensional hydrodynamical simulations.}


   \maketitle

% ###########################################################################################
\section{Introduction}

Hot B subdwarf (sdB) stars are core helium burning objects with effective temperatures $T_{\rm eff} = 20,000 - 40,000$ K and surface gravities $\log g = 5.2-6.2$, belonging to the Extreme Horizontal Branch \citep[EHB;][]{Heber09, Heber16, Heber24}. Their progenitors are low-mass stars ($\sim 1 - 2\Msun$) which, after ascending the Red Giant Branch (RGB), lose most of their envelope near the tip of the RGB (hereafter RGB tip) and ignite helium through a series of flashes. The mass distribution of sdB stars is strongly peaked at $\sim 0.47\Msun$, although extended wings \citep{Schaffenroth22, Latour26} may indicate origin from intermediate-mass progenitors ($\gtrsim 2\Msun$), igniting helium under non-degenerate conditions \citep{Arancibia-Rojas24}. The majority of sdBs is found in binary systems \citep[e.g.][]{Maxted01, Vos18}, and their formation scenarios are dominated by interactions with a stellar companion \citep{Han02, Han03}. In wide binaries, stable Roche-lobe overflow can strip the envelope, while in close systems, common envelope evolution may lead to either envelope ejection or merger, depending on the orbital energy budget \citep{Heber16}. These interactions can leave observable signatures, such as chemical peculiarities or enhanced surface rotation rates \citep{Zhang12, Yu21, Geier11a, Geier13b}.

However, approximately one-third of sdB stars show no evidence of a past or present companion \citep{Copperwheat11, Geier11a}. Single star evolution through enhanced and fine-tuned mass loss at the RGB tip, at the moment of core-He ignition, appears unlikely to account for this population \citep{Dorman93, Sweigart97}. Merger scenarios involving two low-mass helium white dwarfs have been proposed \citep{Iben90, Han03, Zhang12}, but are challenged by several observational constraints, including the low occurrence of compact helium white dwarf binaries \citep{Ratzloff19}, the similar mass distributions of single and binary sdBs \citep{Fontaine12}, and the slow rotation rates of single sdBs, inconsistent with post-merger predictions \citep{Geier12, Charpinet18}. These properties instead suggest a post-RGB origin for single sdBs, akin to red clump stars \citep{Mosser12}. An alternative formation channel is the engulfment of a substellar companion (planet or brown dwarf BD), as first proposed by \citet{Soker98}. While no planetary-mass companions have yet been firmly detected around sdB stars \citep{Thuillier22}, BD companions are known \citep{Schaffenroth14, Schaffenroth15, Schaffenroth21} and may be relatively common \citep{Schaffenroth18}.

The orbital evolution of substellar companions during the main sequence (MS) and RGB phases has been extensively studied \citep[e.g.][hereafter F23]{Villaver09, Villaver14, Privitera16, Fellay23}. Tidal and drag forces, amplified by stellar expansion, drive orbital decay for companions within a few astronomical units (au), leading to engulfment, while more distant companions, in contrast, experience orbital expansion. Engulfed companions are not immediately disrupted; instead, they spiral inward, losing orbital energy and, in some cases, mass, until their final fate is reached \citep{Metzger12, Jia18}. During this inspiral\footnote{In this work we use "inspiral" for the post-engulfment spiralling phase (while the companion revolves within the envelope) and not for the inward migration prior to engulfment.} phase, companions may either survive the interaction, or be destroyed through tidal disruption upon filling their Roche lobe, or via gradual mass loss through thermal or hydrodynamic ablation, although recent studies suggest that tidal disruption is the dominant destruction mechanism \citep{Staff16, OConnor23}.

Engulfment of substellar bodies by giant stars has been explored using both analytical and hydrodynamical (1D-3D) studies \citep[e.g.][]{Nelemans98, Siess99b, Nordhaus06, Staff16, Kramer20, Yarza23, OConnor23}. These studies  show that inspiralling companions can deposit significant orbital energy and angular momentum into the stellar envelope, potentially leading to enhanced mass loss, partial stripping, or complete ejection. However, the minimum companion mass required to eject the envelope remains highly uncertain. While in some cases little of the envelope is removed \citep{Staff16, OConnor23}, \citet{Yarza23} reported that substellar bodies as light as $\approx 10\,\Mj$ can eject the envelope at the tip of the RGB for a 1\,$M_{\odot}$ star. \citet{Kramer20} found in their setup a minimal companion mass of about 0.03\,$M_{\odot}$ ($\approx 30\,\Mj$) for a 0.77\,$M_{\odot}$ star at the RGB tip (originating from 1\,$M_{\odot}$ MS star). These discrepancies partly reflect the fact that these studies focus on a single (or limited) range of stellar and companion masses, making it difficult to disentangle the respective roles of stellar structure, evolutionary stage, and companion mass.

A necessary, although not sufficient, condition for envelope ejection is that the orbital energy released during the inspiral exceeds the envelope binding energy, $E_\bind$. The latter depends strongly on the stellar mass and evolutionary state of the primary, whereas the available orbital energy is determined mainly by the companion mass, the disruption radius, and the stellar radius at the onset of engulfment. The fraction of this energy that can be converted into envelope unbinding remains uncertain and is typically parametrized by the traditional common-envelope efficiency factor $\alpha$ \cite[e.g.][]{Livio1988}. The evolutionary stage of the primary at engulfment is therefore critical. In particular, the formation of an sdB star not only requires the removal of nearly the entire hydrogen-rich envelope ($M_{\rm env} \lesssim 0.02\Msun$ for EHB stars), but also the ignition of core helium burning. Envelope ejection must therefore occur close to the RGB tip, when the stellar core is sufficiently massive to initiate core helium burning \citep{Arancibia-Rojas24}.

In this work, we investigate a broad range of stellar and companion properties and explore the impact of the energy-transfer efficiency on envelope ejection. We consider substellar companions spanning the planetary and brown dwarf regimes ($1-80\,\Mj$) and orbital separations of $0.5-3.5$ au around primary stars with masses between $1.2$ and $2.0\Msun$. By comparing the available inspiral energy, accounting for different deposition efficiencies, with the corresponding envelope binding energy, we identify the conditions required for envelope ejection. Using these constraints and the observed population of exoplanets and brown dwarfs, we estimate the occurrence of envelope ejections through substellar companion engulfment.

This paper is organized as follows. Section \ref{sec:method} describes our orbital evolution code, \texttt{Sekhmet}\footnote{The code is named after the Egyptian goddess Sekhmet whose attributions were, among others, the destruction using the power of the Sun. The code is available on Github: https://github.com/anthuil/Sekhmet}, and the computation of the energy budget. Section \ref{sec: Orbital evolution} presents the orbital evolution of substellar companions during the subgiant and RGB phases. Section \ref{sec: Energetical arguments} outlines the energetic arguments for envelope ejection. Section \ref{sec: Occurrences ejection} estimates the occurrence of envelope ejection events. Section \ref{sdBformation} discusses the additional conditions required for sdB formation. We summarize our results and conclude in Section \ref{sec: Conclusion}.

% #############################################################################################################
\section{Methods}
\label{sec:method}

\subsection{Theoretical framework}
\label{sec:Sekhmet}

To model the orbital evolution of substellar companions around stars evolving through the subgiant and RGB phases, we developed the 1D code \texttt{Sekhmet}. The code accounts for the effects of stellar mass loss, tidal interactions, and drag forces, based on the formalism developed by \citet{Villaver14}, \citet{Bolmont16}, F23 and \citet{OConnor23}. The evolution of the semi-major axis $a$ is given by
	\begin{equation}
		\dot{a} = \dot{a}_{\orb} + \dot{a}_\drag ~,
		\label{eq:dot a general}
	\end{equation}
where $\dot{a}_{\orb}$ accounts for orbital changes due to tidal interactions and stellar mass loss, while $\dot{a}_{\drag}$ represents gravitational and frictional drags. 
The first term in Eq.~\ref{eq:dot a general} is given by 
	\begin{equation}
		\dot{a}_\orb = a \left( 2 \frac{\dot{J}_\orb}{J_{\orb}} - 2 \frac{\dot{M}_\star}{M_\star} + \frac{\dot{M}_\star}{M_\star + M_p} + \frac{2 e \dot{e}}{1-e^2} \right) ~,
		\label{eq:dot a tot}
	\end{equation}
where $J_\orb$ is the orbital angular momentum, $\dot{M}_\star$ the stellar mass-loss rate, $M_\star$ and $M_p$ the stellar and companion masses, respectively, and $\dot{e}$ the rate of change of the eccentricity $e$.
Angular momentum conservation imposes that
	\begin{equation}
		\dot{J}_{\orb} = \dot{J}_{\sys} - \dot{J}_p - \dot{J}_\star ~,
		\label{eq:dot J orb components}
	\end{equation}
where $\dot{J}_{\sys}$ is the rate of angular momentum loss from the system due to the stellar wind, and $\dot{J}_p$ and $\dot{J}_\star$ are the torques on the companion and on the star, respectively. 
Assuming a fast, isotropic stellar wind (Jeans mode), the systemic  angular momentum loss rate is given by 
    \begin{equation}
        \dot{J}_{\sys} = \dot{M}_\star \frac{M_p}{M_\star + M_p} \frac{J_{\orb}}{M_\star} ~.
        \label{eq:dot J sys}
    \end{equation}
Since we consider interactions between the planet/brown dwarf and a post-main sequence star with an extended convective envelope, the tidal torque on the companion, $\dot{J}_p$, results from equilibrium tide and can be expressed as (see Eq.~18 of F23)
	\begin{equation}
		\dot{J}_p = \dot{J}_{\eqp} = J_{\orb} \left( \frac{\dot{a}_{\eqp}}{2 a} - \frac{e~ \dot{e}_{\eqp}}{1-e^2} \right) ~,
		\label{eq:dot J p}
	\end{equation}
where, following F23, $\dot{a}_{\eqp}$ and $\dot{e}_{\eqp}$ are given by	
	\begin{eqnarray}
	    \dot{a}_{\eqp} & = & 6 a k_{2,p} \Delta t_p \frac{1}{q} \left( \frac{R_p}{a} \right)^5 n^2 \left( f_2 \frac{\Omega_p}{n} - f_1 \right) ~, 	\label{eq:dot a eq p} \\
		\dot{e}_{\eqp} & = & \frac{33}{2} e k_{2,p} \Delta t_p \frac{1}{q} \left( \frac{R_p}{a} \right)^5 n^2 \left( f_4 \frac{\Omega_p}{n} - \frac{18}{11} f_3 \right) ~.
        \label{eq:dot e eq planet}
	\end{eqnarray}
Here $k_{2,p}$ is the companion's potential Love number of degree 2, $\Delta t_p$ is the tidal time lag, $q=M_p/M_\star$ is the mass ratio, $R_p$ is the companion radius, $n$ the orbital frequency, and $\Omega_p$ the companion's angular velocity. $f_i$ are the eccentricity-dependent frequency components given by \cite{Hut81}.

The stellar torque, $\dot{J}_\star$, includes contributions from equilibrium tides, $\dot{J}_{\eqs}$, and wind angular momentum loss, $\dot{J}_{\wind}$
	\begin{equation}
		\dot{J}_\star = \dot{J}_{\eqs} + \dot{J}_{\wind} ~,
		\label{eq:dot J star}
	\end{equation}
with $\dot{J}_{\eqs}$ analogous to $\dot{J}_p$
	\begin{equation}
		\dot{J}_{\eqs} = J_{\orb} \left( \frac{\dot{a}_{\eqs}}{2 a} - \frac{e~ \dot{e}_{\eqs}}{1-e^2} \right) ~,
		\label{eq:dot J eq star} 
	\end{equation}
and
	\begin{eqnarray}
		\dot{a}_{\eqs} & = & a \frac{f_{\orb}}{\tau} \frac{M_{\env}}{M_\star} q (1+q) \left( \frac{R_\star}{a} \right)^{8} \left( f_2 \frac{\Omega_\star}{n} - f_1 \right) ~, \label{eq:dot a eq star} \\
		\dot{e}_{\eqs} & = & \frac{11}{4} e \frac{f_{\orb}}{\tau} \frac{M_{\env}}{M_\star} q (1+q) \left( \frac{R_\star}{a} \right)^{8} \left( f_4 \frac{\Omega_\star}{n} - \frac{18}{11} f_3 \right) ~.
		\label{eq:dot e eq star}
	\end{eqnarray}
In the last equations, the factor $f_{\orb}$ is defined as $1$ if $\tau < \frac{P_{\orb}}{2}$ and $\left(\frac{P_{\orb}}{2 \tau}\right)^2$ otherwise (F23). Here $M_{\env}$ is the stellar envelope mass, $R_\star$ the stellar radius, $\Omega_\star$ its rotation rate and $P_{\orb}$ the orbital period, given by Kepler's third law \citep{Kepler1619}. The eddy turnover timescale $\tau$ is approximated as \citep{Rasio96, Villaver09}
	\begin{equation}
		\tau = \left( \frac{M_{\env} (R_\star - R_{\env})^2 }{3 L_\star} \right)^{ \frac{1}{3} } ~,
		\label{eq:tau}
	\end{equation}
where $L_\star$ is the stellar luminosity and $R_{\env}$ the radius at the bottom of the convective envelope. Assuming isotropic wind emission, the torque due to stellar mass loss is \citep{Siess13}
	\begin{equation}
		\dot{J}_{\wind} = - \frac{2}{3} \dot{M}_\star \Omega_\star R_\star^2 ~.
		\label{eq:dot J wind}
	\end{equation}
The eccentricity term $\dot{e}$ entering Eq.~\ref{eq:dot a general} includes contributions from both the star and the companion 
	\begin{equation}
		\dot{e} = \dot{e}_{\eqs} + \dot{e}_{\eqp} ~.
		\label{eq:dot e}
	\end{equation}
The last component to consider is the effect of the drag force on the evolution of the separation. Its expression is given by \citep{Villaver14}
    \begin{equation}
		\dot{a}_\drag =- \frac {2 a}{M_p V_{\orb}} (F_f + F_g) ~,
		\label{eq:dot a D}
	\end{equation}
where $V_{\orb}$ is the orbital velocity of the companion, and $F_f$ is the frictional drag arising from relative motion between the stellar wind and the companion 
	\begin{equation}
		F_f = \frac{1}{2} C_d\,\rho V_{\orb}^2 \pi R_p^2 ~,
		\label{eq:F f}
	\end{equation}
where $C_d$ is the dimensionless drag coefficient for a sphere, set to 0.9 \citep{Villaver14}. Assuming a spherical and homogeneous wind, the density of the circumstellar environment writes 
    \begin{equation}
        \rho = \frac{\dot{M}_\star}{4 \pi a^2 V_{\wind}} ~,
        \label{eq:rho}
    \end{equation}
where $V_{\wind}$ is the wind velocity. The gravitational drag $F_g$ is given by
	\begin{equation}
		F_g = 4 \pi \frac{(G M_p)^2}{ c^2_s } \rho \mathcal{M} ~,
		\label{eq:F g}
	\end{equation}			
where $\mathcal{M}$ is the Mach number, set to 0.5 \citep{Villaver09, Ostriker99}. $c_s$ is the sound speed in the circumstellar environment
	\begin{equation}
		c_s = \sqrt{\frac{5}{3} \frac{\mathcal{R} T_{\wind}}{\mu}} ~,
		\label{eq:C s}
	\end{equation}
where $\mathcal{R}$ is the ideal gas constant, $T_{\wind}$ the wind temperature and $\mu$ the mean molecular weight. For a solar composition \citep{Asplund09} and atomic gas, $\mu = 1.38$.

The evolution of the companion's angular velocity is given by
	\begin{equation}
		\dot{\Omega}_p = \frac{\dot{J}_p} {I_p} ~,
		\label{eq:dot Omega p}
	\end{equation}
where the moment of inertia $I_p$ is assumed to be constant.  The evolution of the stellar angular velocity accounts for structural changes in the star and writes
	\begin{equation}
		\dot{\Omega}_\star = \frac{\dot{J}_\star I_\star - J_\star \dot{I}_\star} {I_\star^2} ~.
		\label{eq:dot Omega star}
	\end{equation}
where  $I_\star$ is the star's moment of inertia.
All stellar quantities needed  for the calculations ($L_\star$, $R_\star$, $I_\star$, etc.) are provided by 1D stellar evolution models computed with the \texttt{STAREVOL} code \citep{Siess00, Siess06}. We emphasize that the orbital evolution during the inspiral phase within the star is not explicitly followed; instead, the disruption location is estimated as described in the next section.
\subsection{Energy balance}
\label{sec:energy balance}

In this section, we describe our calculation of the energy available for envelope ejection and examine the evolution of the envelope binding energy.

\subsubsection{Envelope energy deposition}

In the absence of hydrodynamical simulations, we estimate the energy available for envelope ejection from the change in orbital energy between the onset of engulfment and the point at which the companion is disrupted within the stellar interior. We define engulfment as the point at which the companion reaches the stellar surface ($a = R_\star$), while the endpoint of the inspiral is taken to be the disruption radius, $a_\disrupt$. We do not model the detailed orbital evolution within the envelope, nor do we account for additional processes that may contribute to its unbinding, such as ionization and recombination energy \citep{Lau22a,Lau22b} or radiative losses from the stellar surface. Instead, we adopt the standard common-envelope energy formalism, in which a fraction $\alpha$ of the released orbital energy contributes to the unbinding of the envelope \citep[e.g.,][]{De_Marco11}. The efficiency parameter $\alpha$ has been constrained through reconstructions of post-common-envelope binary evolution. For systems involving brown dwarf companions to main-sequence stars or white dwarfs, the studies of \citet{Zorotovic10, De_Marco11, Zorotovic22} infer $\alpha \simeq 0.2-1$, with little evidence for a dependence on companion mass. The envelope ejection is therefore expected when
\begin{equation}
    |E_\bind| < \alpha E_\orb.
\end{equation}
The orbital energy at disruption is estimated as \citep{OConnor23}
	\begin{equation}
		E_\orb \approx \frac{G M_{\mathrm{int}} M_p}{2 a_{\disrupt}} ~,
		\label{eq:E-orb}
	\end{equation}
where $M_{\mathrm{int}}$ is the mass enclosed within the disruption radius $a_{\disrupt}$, defined as
    \begin{equation}
        a_{\disrupt} = \max(a_{\Roche}, a_{\mathrm{virial}}) ~,
        \label{eq:a disrupt}
    \end{equation}
where $a_{\rm Roche}$ is the Roche radius, corresponding to the radius at which tidal forces overcome the self-gravity of the companion. It is given by \citep{OConnor23}
    \begin{equation}
		a_{\Roche} \simeq 2 R_p \left( \frac{M_{\mathrm{int}}}{M_p} \right)^{\frac{1}{3}} = 2 \left( \frac{M_{\mathrm{int}}}{\frac{4}{3} \pi \bar{\rho}_p} \right)^{\frac{1}{3}} ~,
		\label{eq:R Roche}
	\end{equation}
where $\bar{\rho}_p$ is the mean density of the companion. The virial radius $a_{\mathrm{virial}}$ is defined as the location within the star where the local temperature reaches the companion's virial temperature, given by \citep{Siess99a, Siess99b}
    \begin{equation}
		T_{\mathrm{virial}} \approx 2.4 \times 10^5\, \frac{M_p}{\Mj} \left( \frac{R_p}{\Rj} \right) ^{-1} ~~ [K]~.
		\label{eq: T virial}
	\end{equation}
When the companion temperature globally exceeds $T_{\mathrm{virial}}$, its internal energy becomes higher than its potential energy, leading to its evaporation.

Using our \texttt{STAREVOL} models, we track the radial and mass-coordinate locations of isotherms corresponding to temperatures of 0.5, 1, 2, 3, 5, 10, and $20\times10^6$ K. We define the minimum orbital radius reached before evaporation, $a_{\rm virial}$, as the location of the isotherm whose temperature first exceeds the companion's virial temperature. In our calculations, we neglect ablation and accretion, such that the companion mass remains constant until it reaches $a_\disrupt$, where it is assumed to be instantaneously destroyed (see Sect 3.3 of \cite{OConnor23} for a discussion of this approximation).

\subsubsection{Binding energy of the envelope}

The binding energy of the envelope $E_\bind$ is defined as
	\begin{equation}
		E_\bind = \int_{M_\mathrm{int}}^{M_\star} \left( -\frac{G~M_r}{r} + u \right) dM_r ~,
		\label{eq:E bind}
	\end{equation}
where $M_r$ is the mass enclosed within radius $r$ and $u$ the specific internal energy. By convention, $E_\bind$ is negative for gravitationally bound envelope. For comparison with the orbital energy deposited, we will consider the absolute value $|E_\bind|$.

\begin{figure}[h!]
		\centering
	\includegraphics[width=0.49\textwidth]{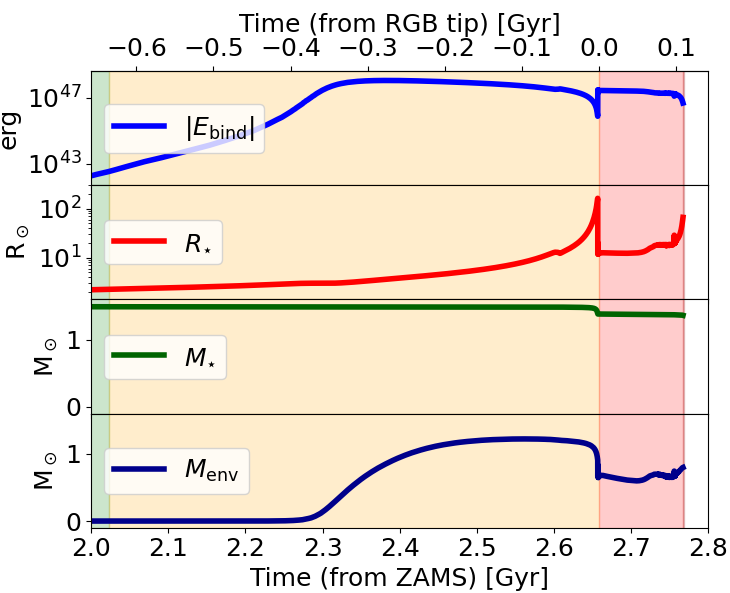}
		\caption{Evolution of the envelope binding energy (top), stellar radius, total and envelope mass for a $1.5\Msun$ star, from the end of the main sequence to the end of core helium burning. Colours indicate the evolutionary phases: end of MS (green), subgiant and RGB phases (orange), and horizontal branch (red). The absolute value $|E_\bind|$ is shown.}
		\label{fig:E bind 1}
	\end{figure}

The evolution of the envelope binding energy is computed from the \texttt{STAREVOL} models and shown in Fig.~\ref{fig:E bind 1} for a $1.5\Msun$ star, from the end of the MS (green) through the subgiant and RGB phases (orange), to the Horizontal Branch (core-He burning phase) (red). The absolute value $|E_\bind|$ reaches a maximum during the subgiant phase, and gradually decreases as the star ascends the RGB, due to the expansion of the envelope. Close to the RGB tip, a sharp drop occurs as the stellar radius increases rapidly and during core helium burning, the star is more compact and its envelope more tightly bound. During the subgiant and RGB evolution, the stellar mass decreases steadily due to wind mass loss, with the strongest reduction occurring near the RGB tip. The envelope mass, $M_\mathrm{env}$, initially increases during the subgiant phase and reaches a maximum at the first dredge-up. It subsequently decreases as the stellar core grows and mass is removed through stellar winds.

We computed \texttt{STAREVOL} evolutionary tracks for stellar masses of $1.0$, $1.2$, $1.5$, $1.7$, and $2.0\Msun$. We use the \cite{Asplund09} solar composition, a mixing length parameter $\alpha_\mathrm{MLT} = 1.75$ and the \cite{Schroeder05} mass loss rate. No core overshooting has been considered. Details about the default input physics (nuclear network, opacities, equation of state,...) can be found in \cite{Siess00} and \cite{Siess06}. As shown in Fig.~\ref{fig:E bind multistar}, the envelope binding energy, $|E_\bind|$, generally increases with stellar mass. In the $1.0\Msun$ model, $E_\bind$ becomes positive near the RGB tip, suggesting that the envelope could be ejected without additional energy input and potentially result in hot subdwarf formation \citep{Han13}. Since this spontaneous envelope ejection scenario remains uncertain, we exclude the $1.0\Msun$ model from the subsequent analysis.

% #################################################################################################################
\section{Orbital evolution}
\label{sec: Orbital evolution}

In this section, we analyse the influence of main parameters on the orbital evolution of the substellar companion (planet or BD) during the subgiant and RGB phases. Our reference simulation considers a 1.5\,$M_{\odot}$ star. The companion is initially placed on a circular orbit with semi-major axis $a = 2$ au and has the mass and radius of Jupiter. The stellar rotation rate $\Omega_\star$ is set to $10^{-7}$ rad/s, typical of RGB stars \citep{Ceillier17}, while the companion angular velocity $\Omega_p$, is set to $7 \times 10^{-5}$ rad/s (rotation period of $\sim$ 1 day). The tidal parameter is fixed to $k_{2,p} \Delta t_p=0.01$ s (F23), and the moment of inertia factor of the companion is $C_{I, p} = 0.2$, consistent with massive planets or BDs \citep{Chabrier00}. The wind properties are set to a temperature of 2500\,K, a mean molecular weight $\mu = 1.38$, and a velocity $V_{\rm wind} = 5~\mathrm{km\,s^{-1}}$. The simulation starts 2.56 Gyr after the ZAMS and stops either at engulfment or, if no engulfment occurs, 7 Myr after the RGB tip when the star has passed the He flash. During the integration, the maximum relative variation of $a$, $e$, and $\Omega_\star$ between consecutive time steps is limited to 1\%. Unless otherwise specified, these parameters define the reference case used throughout this section.

% ----------------------------------------------------------------------------------------------------------------
\subsection{Major contributions to {$\dot{a}$}}

\begin{figure}[h]
		\centering
		\includegraphics[width=0.49\textwidth]{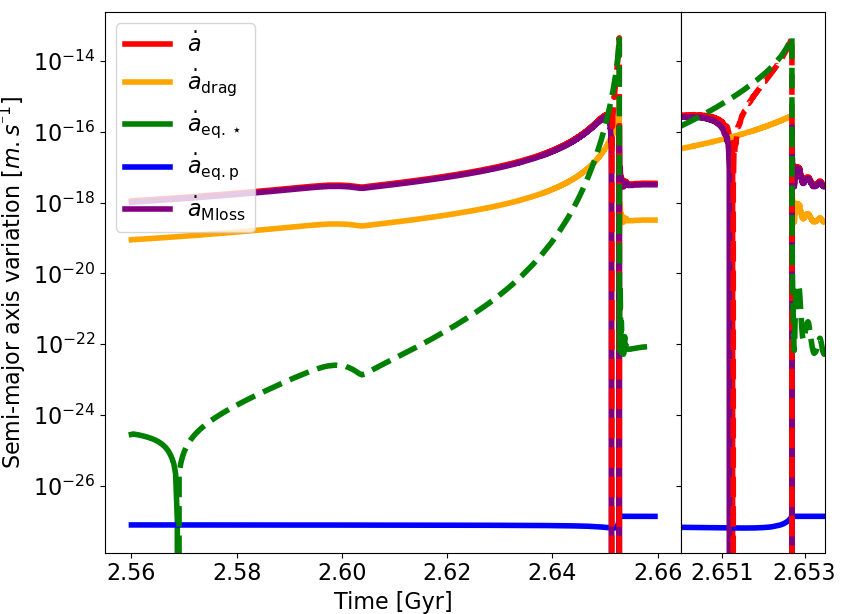}
		\caption{Contributions to the rate of change of the semi-major axis $\dot{a}$ (solid red line) in the  $1.5\Msun$ reference model with a Jupiter companion initially located at 2 au: drag forces (orange), companion (blue) and stellar (green) equilibrium tides, and stellar mass loss (violet). The total $\dot{a}$ is shown as a solid red line. Positive (negative) values correspond to orbital expansion (decay). Negative contributions are shown as dashed lines, plotted as $\log(-\dot{a})$. The right panel zooms in on the RGB tip.}
		\label{fig:a components}
	\end{figure}

Figure~\ref{fig:a components} shows the individual contributions to the evolution of the semi-major axis (red curve) in the reference case. During the post-MS evolution, stellar mass loss ($\dot{a}_\mathrm{Mloss} = \dot{a}-\dot{a}_{\eqs}-\dot{a}_{\eqp}$, violet curve) dominates the orbital evolution, with drag forces (orange curve) contributing at the $\sim 10\%$ level in our reference model. Both terms are positive and lead to a gradual increase of the orbital separation. Near the RGB tip, the rapid expansion of the stellar envelope strongly enhances tidal interactions. Stellar equilibrium tides (green curve) then dominate the evolution, exceeding the contribution from mass loss by up to two orders of magnitude, and act to decrease the orbital separation. In contrast, equilibrium tides raised on the companion (blue curve) remain negligible throughout the evolution. During core helium burning, the star is more compact and the tidal effect very weak, which explains the plateau in Fig.~\ref{fig:Infl. a-init}.

% ----------------------------------------------------------------------------------------------------------------
\subsection{Dependence on the initial semi-major axis}

\begin{figure}[h!]
		\centering
		\includegraphics[width=0.49\textwidth]{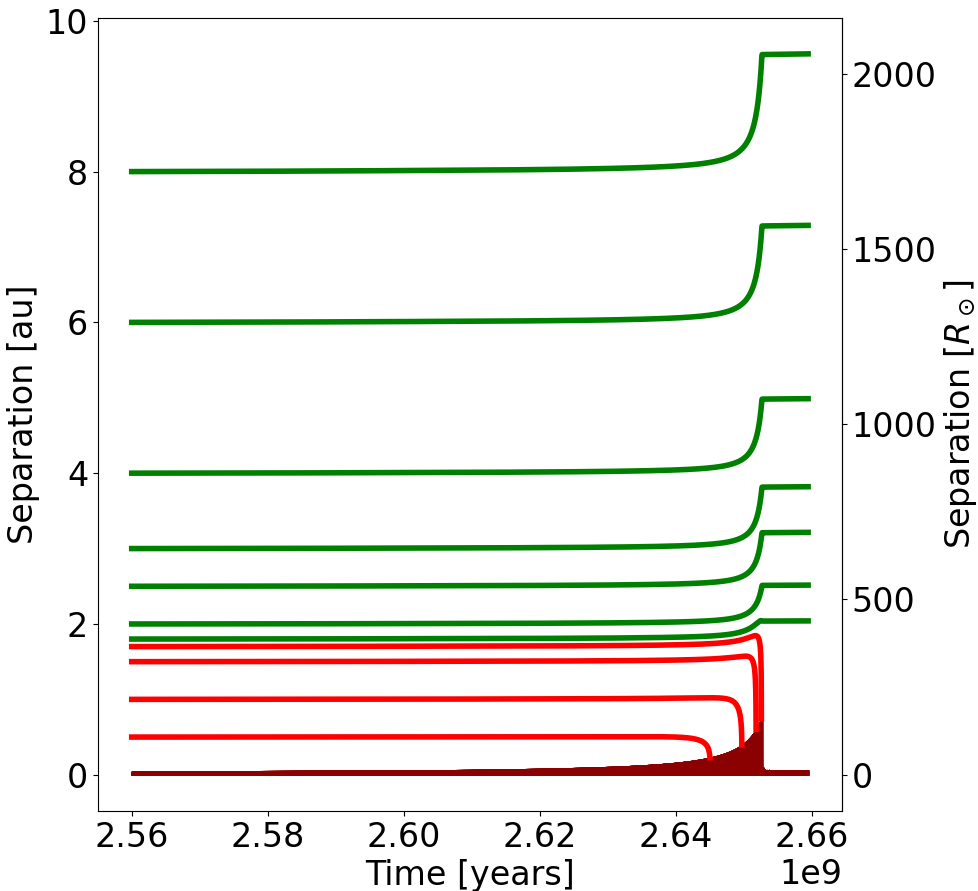}
		\caption{Evolution of the semi-major axis for companions with initial orbital separation (in \texttt{au}): 0.5, 1, 1.5, 1.7, 1.8, 2, 2.5, 3, 4, 6, 8. Green lines indicate surviving planets, red lines, engulfed ones. The dark red area represents the stellar radius.}
		\label{fig:Infl. a-init}
	\end{figure}

Figure \ref{fig:Infl. a-init} shows the orbital evolution for companions with different initial semi-major axes during the late RGB phase. At early times, orbital expansion is driven by stellar mass loss but as the star ascends the RGB and its radius increases, tidal forces strengthen and eventually dominate the orbital evolution. This leads to a bifurcation in behaviour: companions initially located within a critical orbital separation undergo orbital decay and are engulfed, whereas more distant companions continue to migrate outward and avoid engulfment. Our results are consistent with those of \cite{Villaver09, Villaver14} placing the critical radius around 2 au for a $1.5\Msun$ star with a $1 \Mj$ planet.

% ----------------------------------------------------------------------------------------------------------------
\subsection{Effects of other parameters}

We explored the influence of several parameters on the orbital evolution, including orbital eccentricity, stellar and companion rotation rates, mass-loss prescriptions, companion radius, and wind properties (velocity and temperature). Although variations in individual parameters generally produce modest changes, their combined effects may significantly modify the orbital evolution history. Among these factors, eccentricity is expected to play a particularly important role, as it enhances tidal dissipation during close periastron passages and increases the likelihood of engulfment. The stellar mass also affects the efficiency of tidal orbital decay. For example, lower mass stars reach larger radii during the RGB phase (see Table~\ref{tab:Rmax_RGB}), which strengthens tidal interactions and can accelerate the reduction of the companion's orbital separation. Stellar metallicity, which is not considered in this study, may introduce an additional dependence. Metal-rich stars attain larger radii near the RGB tip, and are observed to host massive planets more frequently \citep{Gonzalez97, Santos01, Fischer05, Jones16}.

% #############################################################################################################
\section{Envelope ejection}
\label{sec: Energetical arguments}

% -------------------------------------------------------------------------------------------------------------
\subsection{Evolution of the disruption radius}

\begin{figure}[h!]
		\centering
		\includegraphics[width=0.49\textwidth]{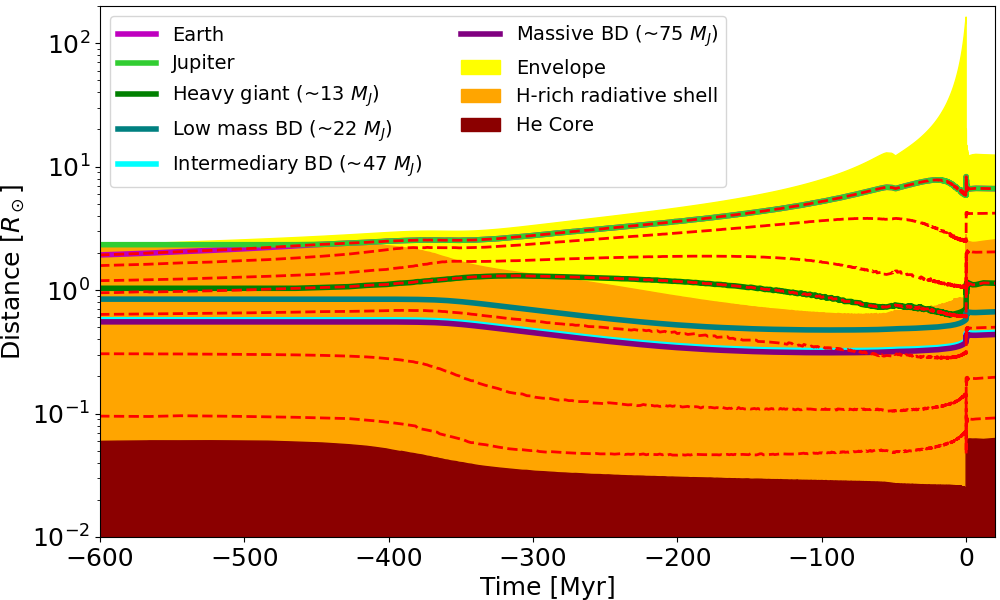}
		\caption{Disruption radius of companions of different masses, assuming engulfment at different evolutionary stages along the RGB of a $1.5\Msun$ star. Red dashed lines indicate isotherms at $0.5,\ 1,\ 2,\ 3,\ 5,\ 10$ and $20 \times 10^6$~K (from top to bottom). Time is measured relative to the RGB tip (t=0).}
		\label{fig:Roche multibody}
	\end{figure}

Figure \ref{fig:Roche multibody} shows the evolution of the disruption radius for seven representative substellar bodies (from Earth-like planets to massive brown dwarfs) as their $1.5\Msun$ host star ascends the RGB. We assume the chemical composition of the companion is homogeneous. In reality, differentiated bodies are expected to follow a more complex evolution: ablative processes may strip the mantle, allowing the denser core to continue its inward migration until it reaches its own disruption radius. 

The Roche radius remains nearly constant during the subgiant phase, decreases as the convective envelope deepens, and increases again during the upper RGB as the core mass grows. For a $1.5\Msun$ star, $a_\Roche$ reaches a minimum of approximately $\sim 100$\,Myr before the RGB tip. At this stage, companions reaching their Roche limit are disrupted deeper within the stellar potential well, resulting in a larger release of orbital energy during inspiral. However, the envelope is still relatively tightly bound, and the impact on envelope ejection depends on the balance between the available orbital energy and the envelope binding energy. At later evolutionary stages, the disruption radius increases slightly while the envelope binding energy decreases sharply as the star expands, modifying the conditions for envelope ejection.

The positions of the 0.5, 1, 2, 3, 5, 10 and 20$\times10^6$ K isotherms broadly track the evolution of the core, with the exception of the outer, lower-temperature curves, which are more sensitive to variations in the stellar radius (see Fig.~\ref{fig:Roche multibody}).

Both $a_{\rm Roche}$ and $a_{\rm virial}$ depend on the density of the engulfed body (see Eqs.~\ref{eq:R Roche}-\ref{eq: T virial}). Denser bodies penetrate deeper into the stellar interior and therefore release more orbital energy into the envelope. In our models, the densest companions are disrupted within the region surrounding the hydrogen-burning shell, below the base of the convective envelope. Brown-dwarf companions are disrupted predominantly at their Roche limits, whereas the disruption radius of planetary-mass companions is more sensitive to the evolutionary state of the host star because their lower densities make the virial condition more restrictive. These results are consistent with previous works \citep{Yarza23}.

% ----------------------------------------------------------------------------------------------------------------
\subsection{Time window for envelope ejection}

\begin{figure}[h!]
		\centering
		\includegraphics[width=0.49\textwidth]{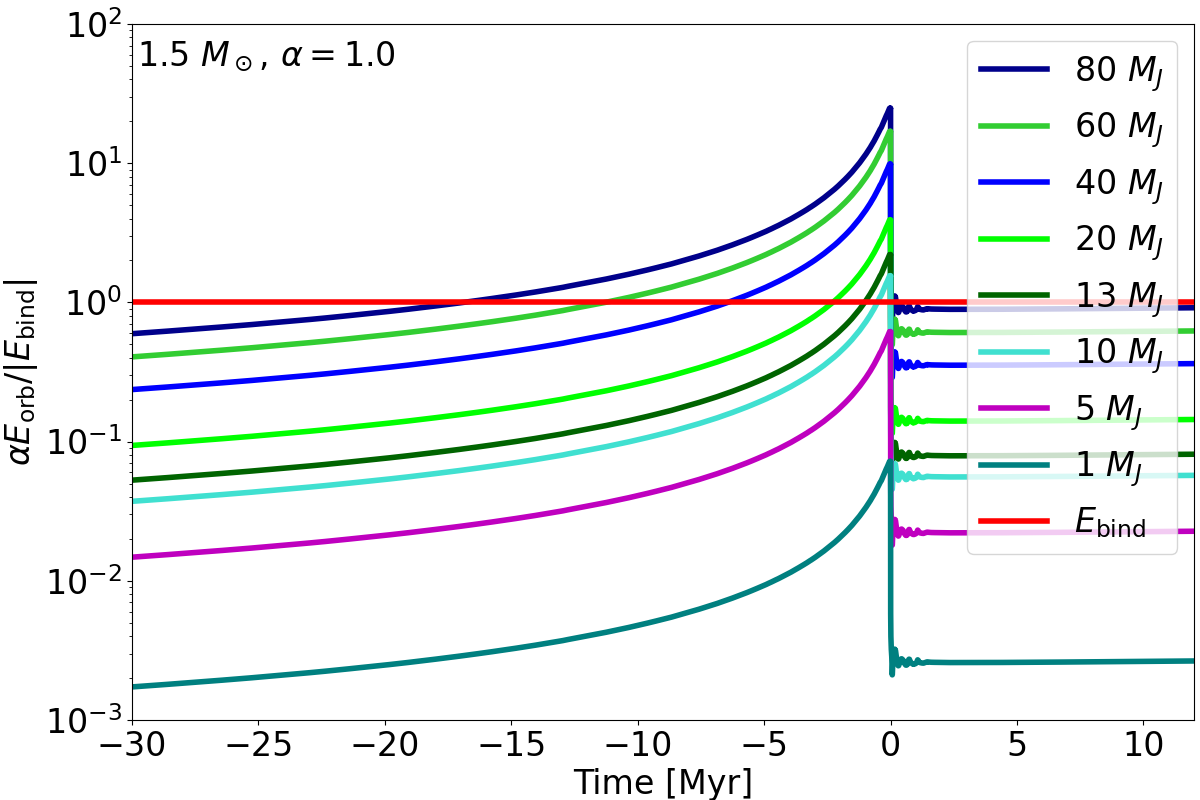}
		\caption{Ratio of the orbital energy to the envelope binding energy in the most favourable case $\alpha=1$  for companions with masses between 1 and $80\,\Mj$ (all assuming $R_{\rm p} = 1\,\Rj$) orbiting a $1.5\Msun$ star. The RGB tip corresponds to time 0 on the abscissa. Values above unity indicate energetically favourable conditions for envelope ejection. Similar graphs for energy transfer efficiencies of 0.2 and 0.5 are shown in Fig.~\ref{fig: E balance allstar}}
		\label{fig:E balance}
	\end{figure}

The most favourable conditions for envelope ejection occur close to the RGB tip, where the envelope binding energy reaches its minimum. Figure \ref{fig:E balance} shows the ratio $E_\orb / |E_\bind|$ in the reference $1.5\Msun$ stellar model for companions with masses in the range $1-80\,\Mj$. In the limiting case of complete energy transfer ($\alpha=1$), a companion mass of at least $\sim 8\,\Mj$ is required for the released orbital energy to equal the envelope binding energy, provided engulfment occurs within the final $\sim 0.2$~Myr before the RGB tip. More massive companions can, in principle, eject the envelope over a substantially longer interval. For example, an $80\,\Mj$ brown dwarf can supply sufficient energy for envelope ejection as early as $\sim 17$~Myr before the RGB tip.

Table~\ref{tab:duration 100pct} and Figure~\ref{fig: E balance allstar} summarize the results for stellar masses between 1.2 and $2.0\Msun$. The minimum companion mass required for envelope ejection increases systematically with stellar mass, reflecting the larger binding energies of more massive RGB envelopes. Conversely, for a given primary mass, increasing the companion mass both increases the available orbital energy and extends the time interval over which envelope ejection is possible. For example, in the case $\alpha=1$, the minimum companion mass increases from $\sim 5\,\Mj$ for a $1.2\Msun$ star to $\sim 20\,\Mj$ for a $2.0\Msun$ star, while the permitted engulfment interval expands from $\sim 0.43$ Myr for a $5\,\Mj$ companion to $\sim 28$ Myr for an $80\,\Mj$ brown dwarf around a $1.2\Msun$ primary. We also find that lower values of $\alpha$ require progressively more massive companions to achieve envelope ejection, since a smaller fraction of the released orbital energy is deposited into the envelope. 

\begin{table}[h!]
		\begin{center}
		\begin{tabular}{|c|c|c|c|c|c|}
		\hline
		$\alpha$ & \text{$M_\mathrm{Companion}$} & \multicolumn{4}{|c|}{\text{Duration with $\alpha E_\orb > |E_\bind|$}}\\
		& {[$\Mj$]} & \multicolumn{4}{|c|}{\text{[Myr]}}\\
		\cline{3-6}
		  & & \text{1.2 $\Msun$} & \text{1.5 $\Msun$} & \text{1.7 $\Msun$} & \text{2.0 $\Msun$}\\
		\hline
		\multirow{8}{1em}{$1$} & 1         &  \cellcolor[HTML]{b0bec5}  --    &   \cellcolor[HTML]{b0bec5} --    &  \cellcolor[HTML]{b0bec5}  --    &  \cellcolor[HTML]{b0bec5}  --   \\
		%3             &    --    &    --    &    --    &    --   \\
		& 5             &  $0.43$ &  \cellcolor[HTML]{b0bec5}  --    &  \cellcolor[HTML]{b0bec5}  --    &  \cellcolor[HTML]{b0bec5}  --   \\
		%7             &  $0.87$ &    --    &    --    &    --   \\
		& 10            &  $1.57$ &  $0.52$ &  \cellcolor[HTML]{b0bec5}  --    &  \cellcolor[HTML]{b0bec5}  --   \\
		& 13            &  $2.30$ &  $1.02$ &  $0.43$ & \cellcolor[HTML]{b0bec5}   --   \\
        \cdashline{2-6}
		& 20            &  $4.15$ &  $2.29$ &  $1.46$ & $0.10$ \\
		& 40            & $10.45$ &  $6.44$ &  $4.78$ & $2.60$ \\
		& 60            & $18.24$ & $11.27$ &  $8.54$ & $5.33$ \\
		& 80            & $27.77$ & $16.90$ & $12.83$ & $8.34$ \\
		\hline
		\hline
        \multirow{8}{1em}{$0.5$} & 1         &  \cellcolor[HTML]{b0bec5}  --    &   \cellcolor[HTML]{b0bec5} -- & \cellcolor[HTML]{b0bec5} -- & \cellcolor[HTML]{b0bec5} -- \\
        & 5 & \cellcolor[HTML]{b0bec5} -- & \cellcolor[HTML]{b0bec5} -- & \cellcolor[HTML]{b0bec5} -- & \cellcolor[HTML]{b0bec5} -- \\
        & 10 & $0.64$ & \cellcolor[HTML]{b0bec5} -- & \cellcolor[HTML]{b0bec5} -- &  \cellcolor[HTML]{b0bec5} -- \\
		& 13 &  $1.05$ &  $0.14$ & \cellcolor[HTML]{b0bec5} -- & \cellcolor[HTML]{b0bec5}   --   \\
        \cdashline{2-6}
		& 20 &  $2.03$ &  $0.84$ &  $0.28$ &  \cellcolor[HTML]{b0bec5} -- \\
		& 40 &  $5.25$ &  $3.01$ &  $2.04$ &  $0.56$ \\
		& 60 &  $8.99$ &  $5.48$ &  $4.02$ &  $2.03$ \\
		& 80 & $13.23$ &  $8.19$ &  $6.15$ &  $3.61$ \\
		\hline
		\hline
        \multirow{8}{1em}{$0.2$} & 1 & \cellcolor[HTML]{b0bec5} -- & \cellcolor[HTML]{b0bec5} -- & \cellcolor[HTML]{b0bec5} -- & \cellcolor[HTML]{b0bec5} -- \\
        & 5 & \cellcolor[HTML]{b0bec5} -- & \cellcolor[HTML]{b0bec5} -- & \cellcolor[HTML]{b0bec5} -- & \cellcolor[HTML]{b0bec5} -- \\
        & 10 &   \cellcolor[HTML]{b0bec5} -- & \cellcolor[HTML]{b0bec5} -- &   \cellcolor[HTML]{b0bec5} -- & \cellcolor[HTML]{b0bec5} -- \\
		& 13 &  $0.20$ & \cellcolor[HTML]{b0bec5} -- & \cellcolor[HTML]{b0bec5} -- &  \cellcolor[HTML]{b0bec5} -- \\
        \cdashline{2-6}
		& 20 &  $0.65$ &  \cellcolor[HTML]{b0bec5} -- &  \cellcolor[HTML]{b0bec5} -- &  \cellcolor[HTML]{b0bec5} -- \\
		& 40 &  $2.04$ &  $0.85$ &  $0.28$ &   \cellcolor[HTML]{b0bec5} -- \\
		& 60 &  $3.59$ &  $1.90$ &  $1.15$ &   \cellcolor[HTML]{b0bec5} -- \\
		& 80 &  $5.29$ &  $3.04$ &  $2.07$ &  $0.57$ \\
		\hline
        
		\end{tabular}
		\end{center}
		\caption{Time interval (in Myr) during which the orbital energy exceeds the envelope binding energy, assuming $\alpha=1$ (top), 0.5 (middle) or 0.2 (bottom). The dashed horizontal line marks the conventional boundary between planets and brown dwarfs at $13\,\Mj$. The grey-shaded areas indicate conditions under which no ejection is possible. For each stellar mass, the minimum companion mass capable of ejecting the envelope corresponds to the first entry in the table. For example, companions with masses equal or below $10\,\Mj$ cannot unbind the envelope of a $1.5\Msun$ star when considering $\alpha =1$.}
		\label{tab:duration 100pct}
	\end{table}

% ----------------------------------------------------------------------------------------------------------------
\subsection{Fate of the substellar companion}
\label{sec: fate of companion}

Figure~\ref{fig:engulf_regions_all_scenarios} shows the possible outcomes in the companion mass--initial semi-major axis plane for different stellar masses and energy-transfer efficiencies. Three regimes are identified: (i) no engulfment, (ii) engulfment without envelope ejection ($\alpha E_\orb<|E_\bind|$), and (iii) envelope ejection ($\alpha E_\orb>|E_\bind|$).

The critical companion mass $M_p^\engl$ that separates engulfed from non-engulfed companions is independent of the energy-transfer efficiency and is confined to relatively large initial separations ($a \gtrsim 2.5$ au). As the stellar mass increases, this boundary shifts toward shorter orbital periods, reflecting the smaller maximum RGB radii reached by more massive stars\footnote{This behaviour is characteristic of low-mass stars ($M\lesssim 2.3\Msun$) that develop degenerate helium cores after the main sequence.} (Tab.~\ref{tab:Rmax_RGB}).
\begin{table}[h!]
\begin{center}
\begin{tabular}{|c|c|c|}
\hline
\text{$M_\star [M_\odot]$} & \text{$R_\mathrm{max}$ $[R_\odot]$} & \text{tip-RGB [Gyr]}\\
\hline
1.2   & $182.60$ & $5.843$ \\
1.5   & $162.60$ & $2.656$ \\
1.7   & $154.13$ & $1.737$ \\
2.0   & $129.55$ & $1.013$ \\
\hline
\end{tabular}
\end{center}
\caption{Radius and age at RGB tip for the \texttt{STAREVOL} model with different initial masses.}
\label{tab:Rmax_RGB}
\end{table}
    
The envelope-ejection region is bounded at low separations by $M_p^\eject$, the minimum companion mass required to unbind the envelope, and at larger separations by $M_p^\engl$. For a given stellar mass and energy-transfer efficiency, $M_p^\eject$ decreases with increasing orbital separation because companions engulfed from wider orbits interact with more extended and less tightly bound envelopes. The minimum separation allowing envelope ejection increases with stellar mass and decreases with increasing $\alpha$, reflecting the larger binding energies of more massive envelopes and the reduced fraction of orbital energy available for unbinding at lower efficiencies. Consequently, the region leading to envelope ejection becomes progressively narrower for more massive primaries and smaller values of $\alpha$.

We derive analytical fits to the boundaries separating these regimes (Appendix~\ref{sec:fate zone borders}), allowing the outcome of a substellar companion to be predicted from its initial mass, $M_p$, orbital separation, $a$, and assumed energy-transfer efficiency without performing full orbital integrations. For each stellar mass, the engulfment and envelope-ejection boundaries (in unit of Jupiter mass) are fitted using the following functional forms:
\begin{eqnarray}
	M_p^\engl(a) & = & \alpha' \exp(\beta a + \gamma) + \delta a + \epsilon 	\label{eq:engulf limit} \\
	M_p^\eject(a) & = &\alpha' a^5 + \beta a^4 + \gamma a^3 + \delta a^2 + \epsilon a + \zeta \ .
	\label{eq:eject limit}
\end{eqnarray}
The fitting parameters are listed in Tables \ref{tab:fate_zones_engulf_coef}.1 and \ref{tab:fate_zones_eject_coef}.2.

If envelope ejection ultimately leads to He ignition, the companion may survive and remain in a close orbit around the newly formed sdB star. Several BD companions to sdB stars in very close orbits ($P_{\rm orb} \lesssim 0.1$ d) are known \citep{Schaffenroth14, Schaffenroth15, Schaffenroth21}, providing strong observational support for this formation channel. Yet, no planet or planetary remnants have been firmly detected around sdB stars \citep{Van_Grootel21, Thuillier22}.

% #############################################################################################################
\section{Occurrence of envelope ejection events}
\label{sec: Occurrences ejection}

	\begin{figure*}[h!]
        \begin{tabular}{|c|c|c|}
        \includegraphics[width=0.31\textwidth]{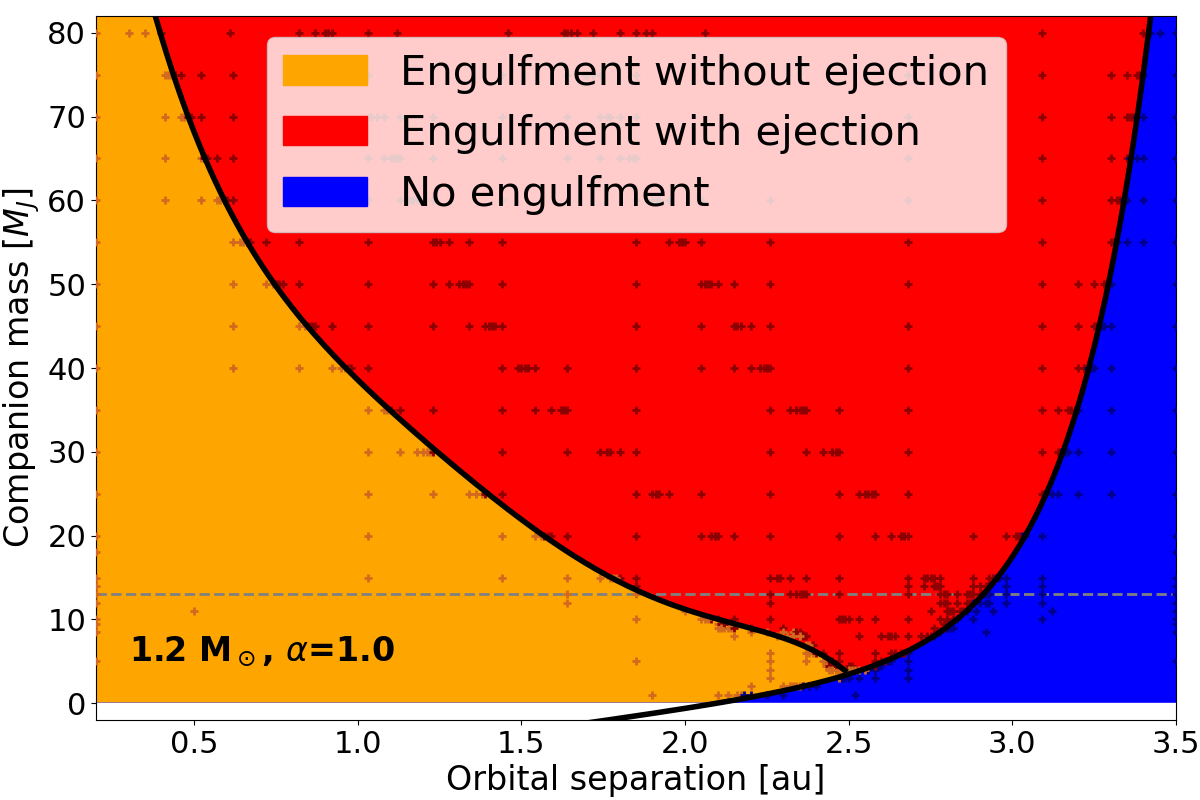} & \includegraphics[width=0.31\textwidth]{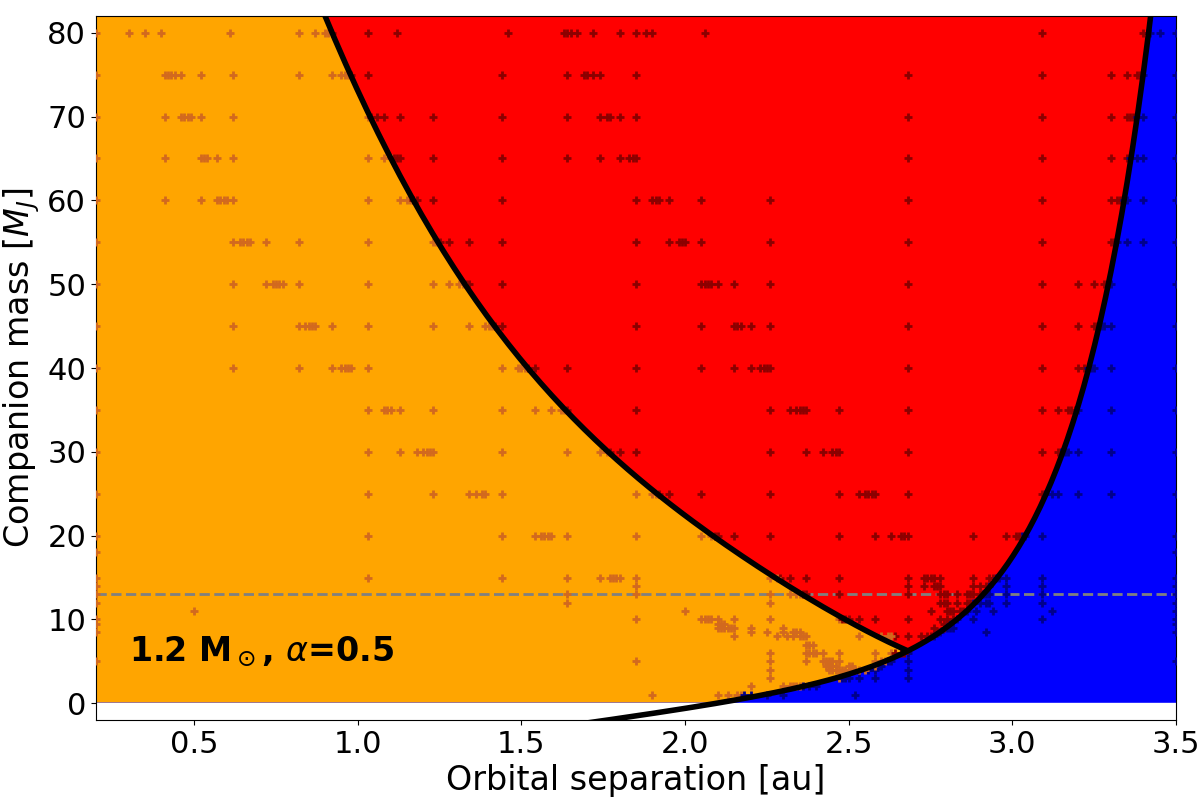} & \includegraphics[width=0.31\textwidth]{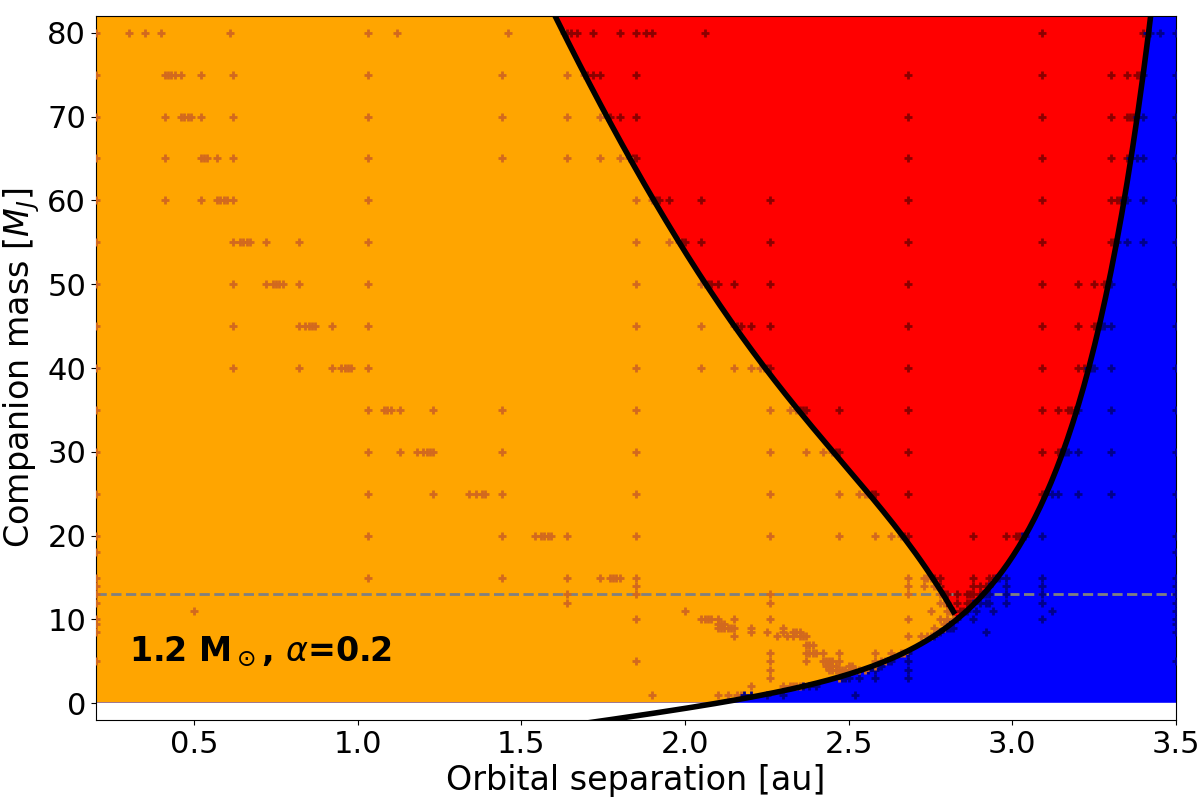} \\
        
        \includegraphics[width=0.31\textwidth]{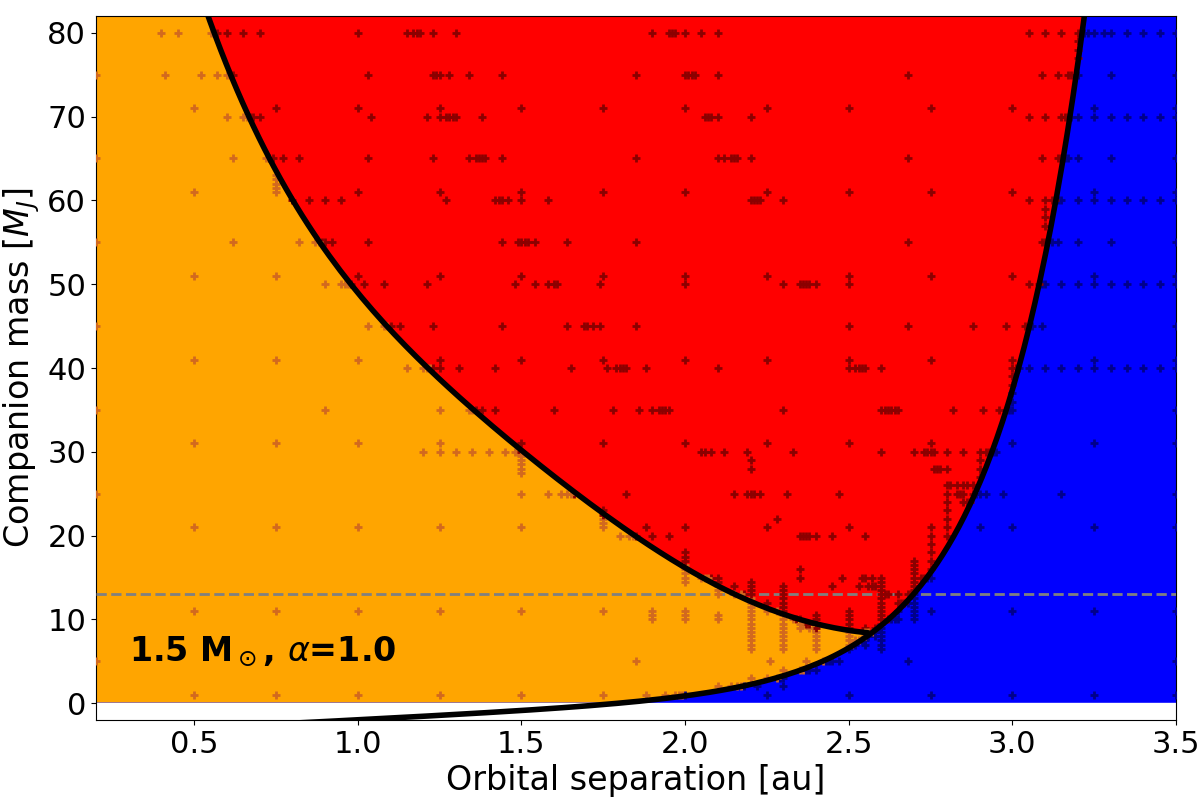} & \includegraphics[width=0.31\textwidth]{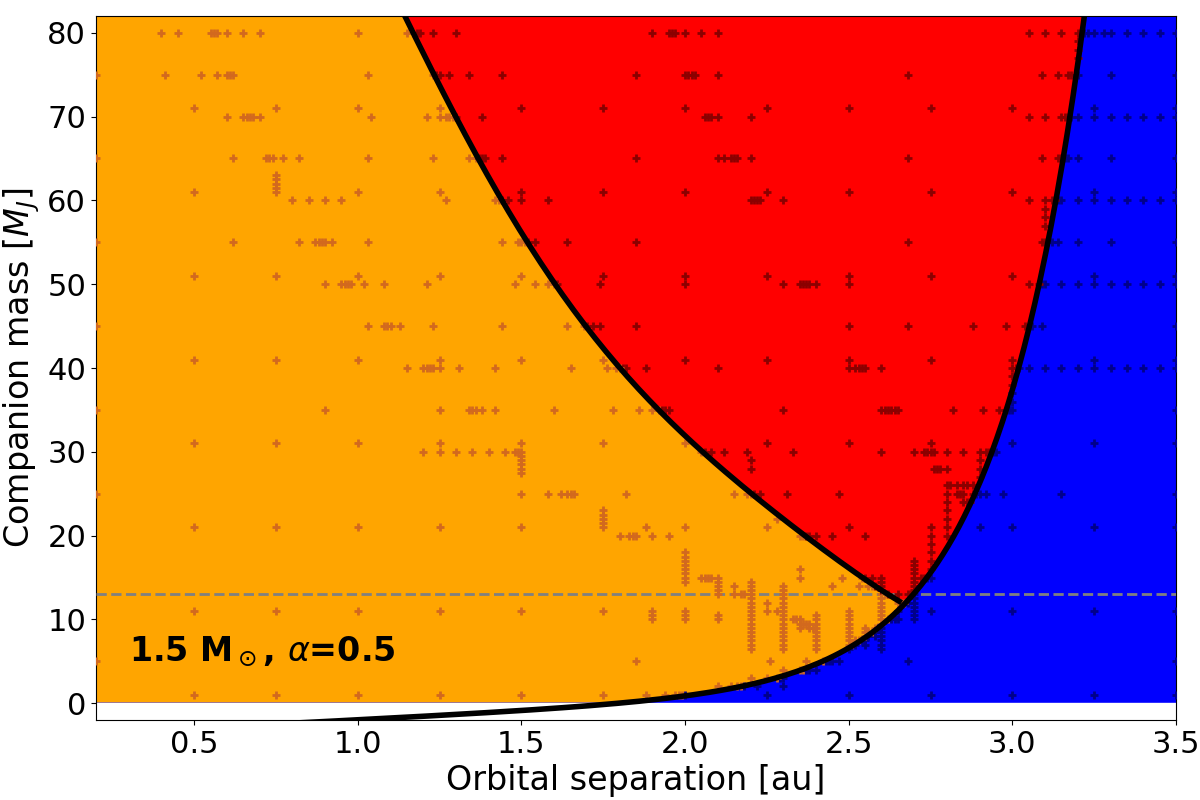} & \includegraphics[width=0.31\textwidth]{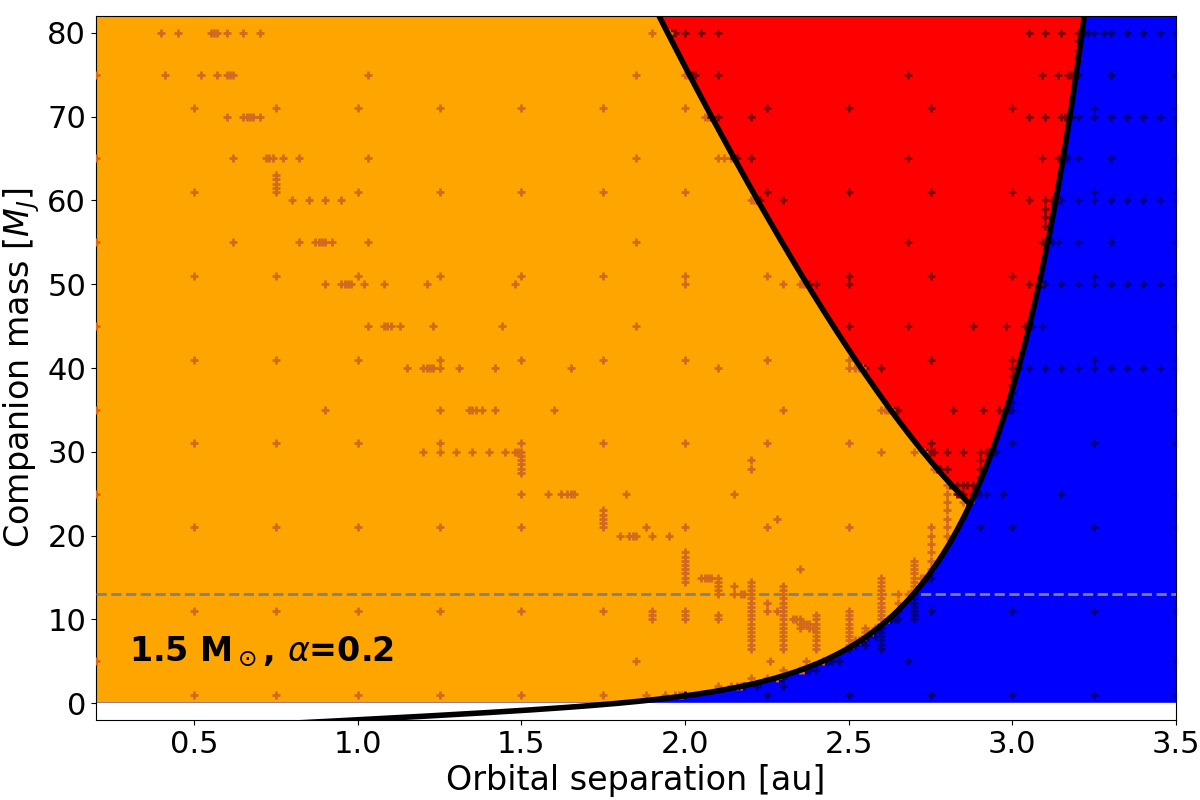} \\
        
        \includegraphics[width=0.31\textwidth]{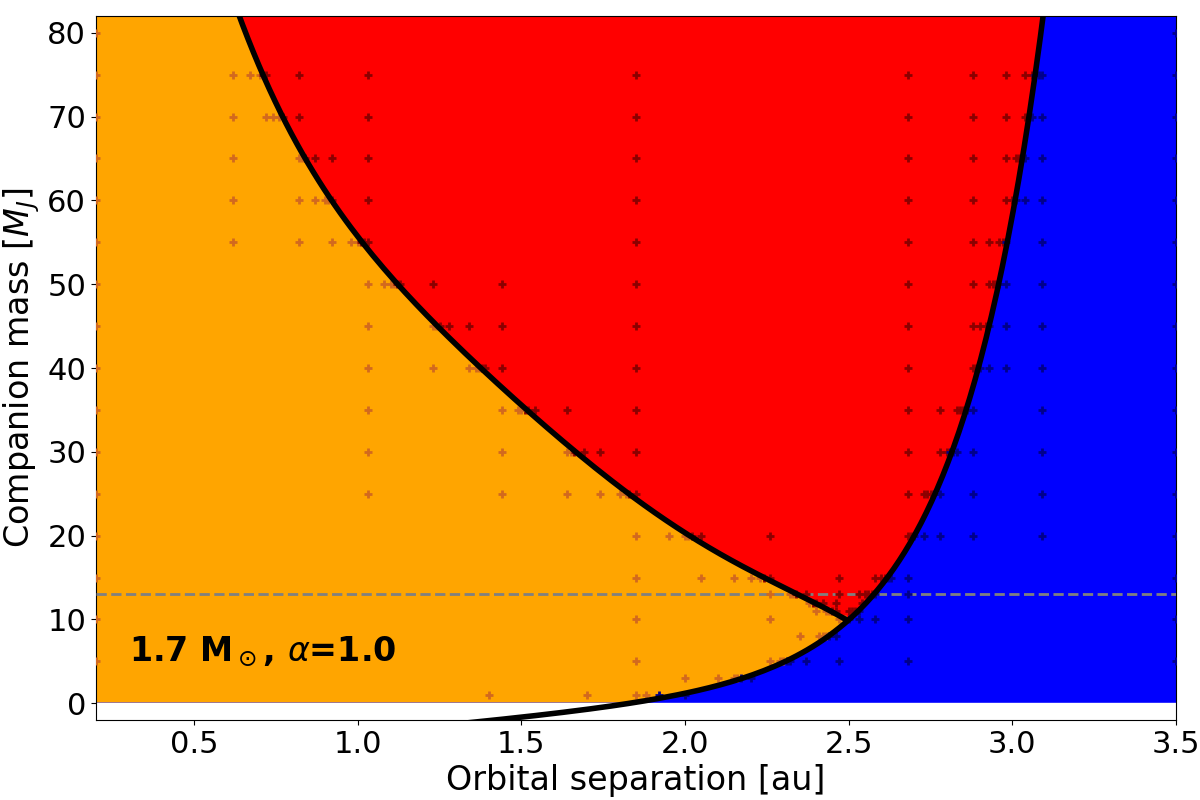} & \includegraphics[width=0.31\textwidth]{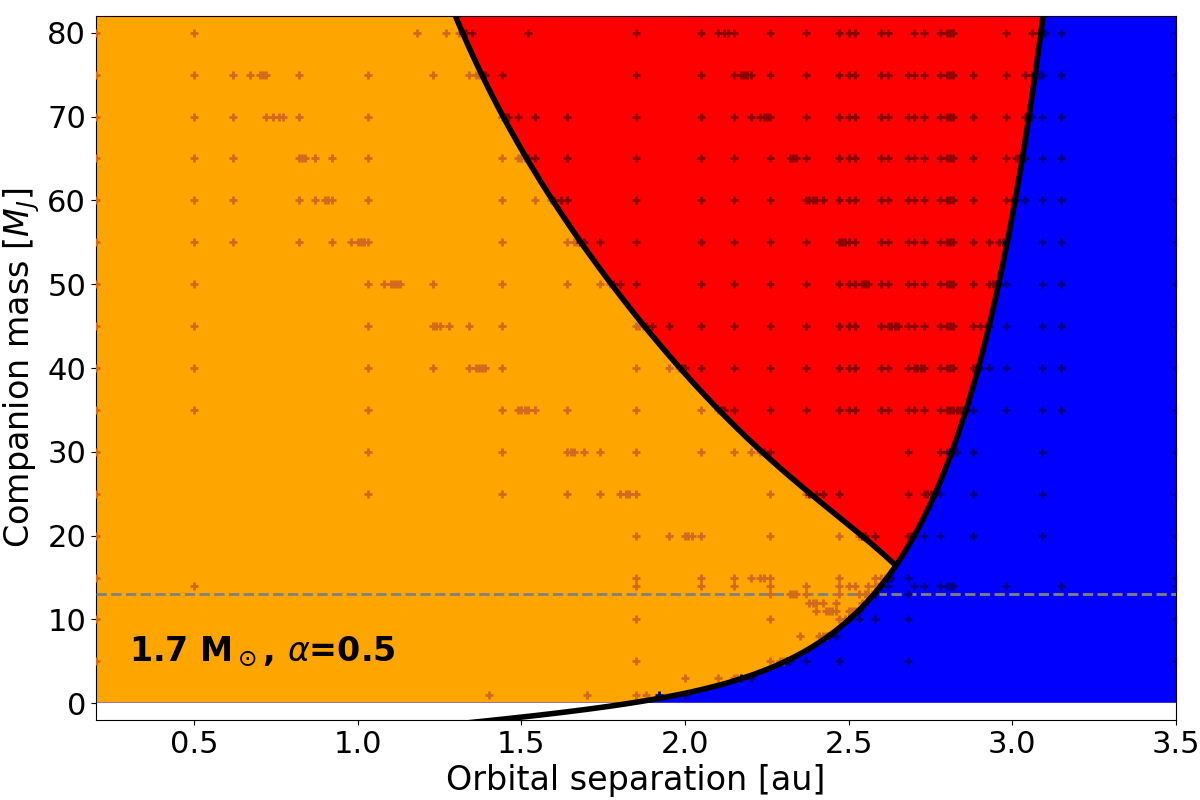} & \includegraphics[width=0.31\textwidth]{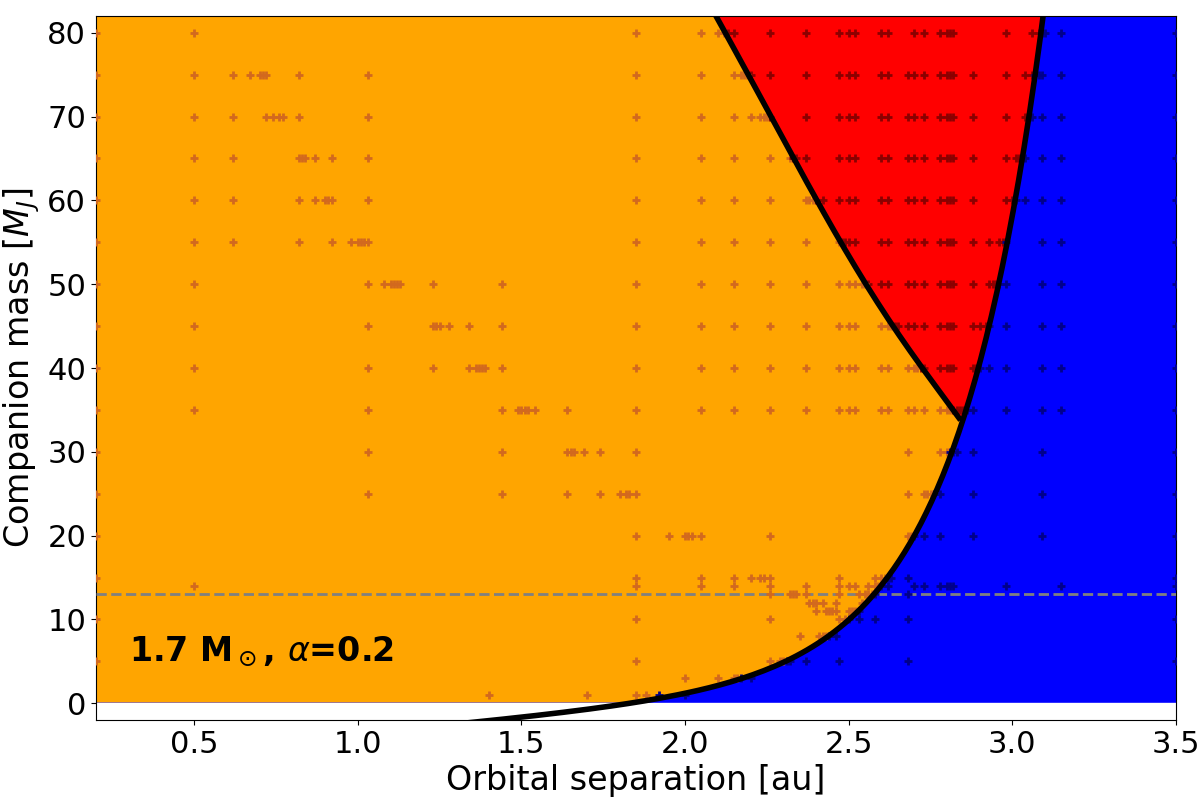} \\
        
        \includegraphics[width=0.31\textwidth]{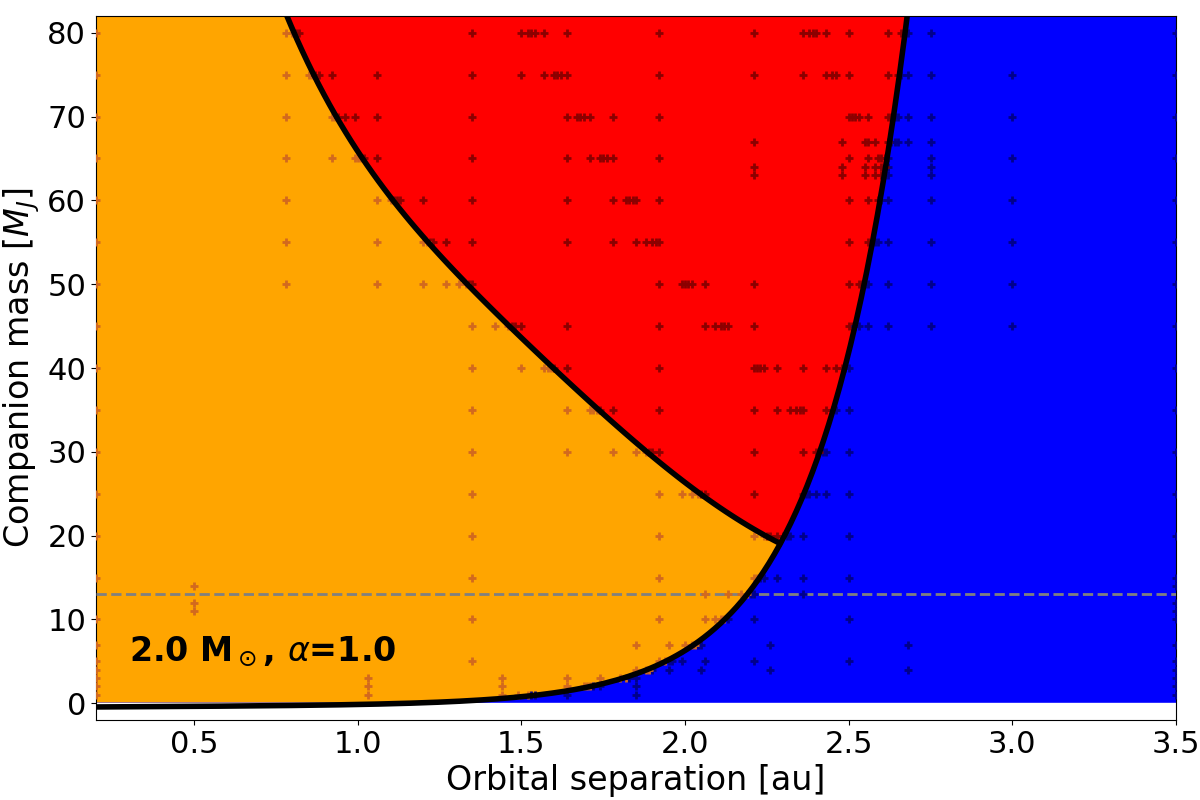} & \includegraphics[width=0.31\textwidth]{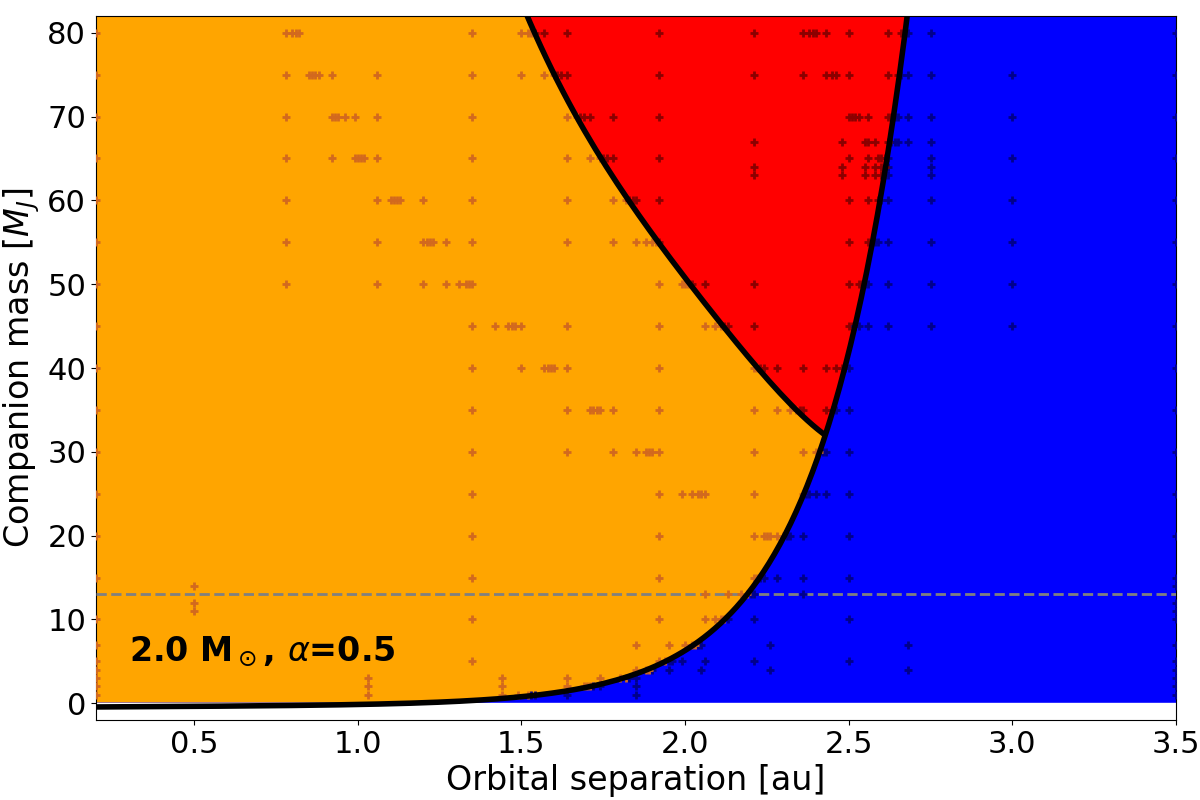} & \includegraphics[width=0.31\textwidth]{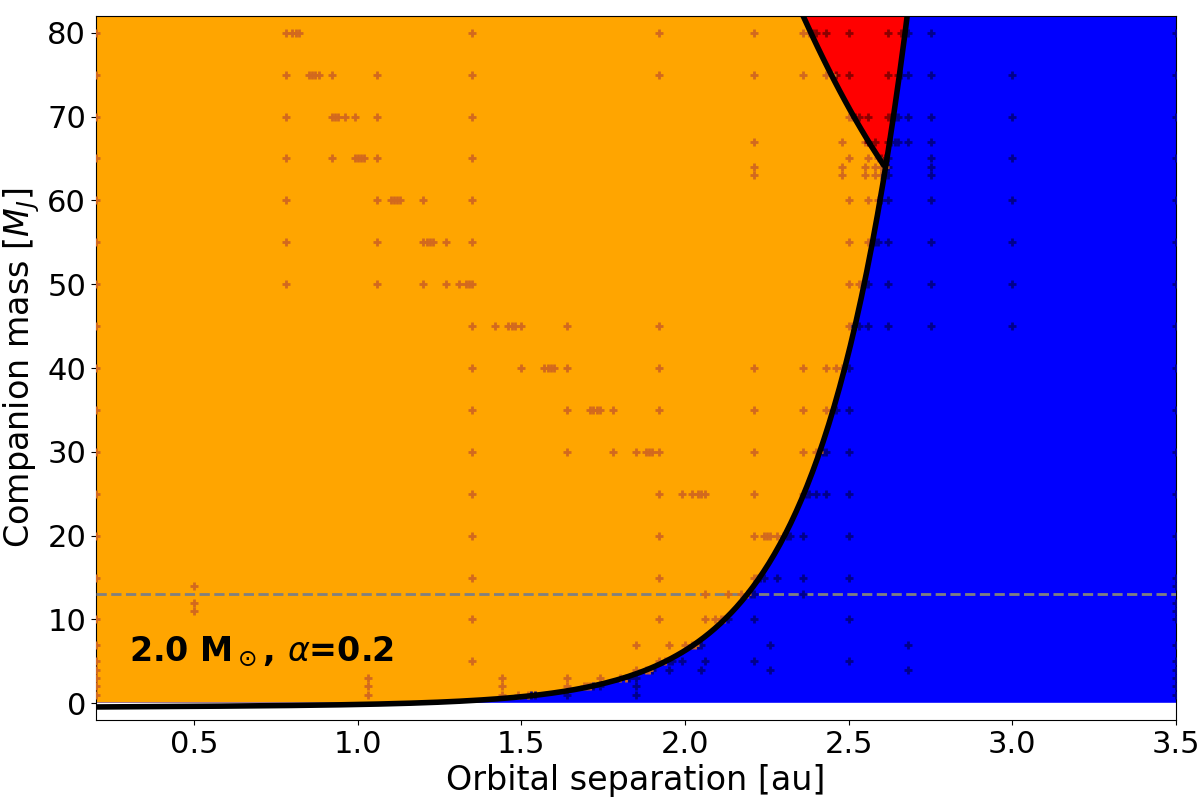} \\
        \end{tabular}
        \caption{Fate of substellar companions as a function of companion mass and initial semi-major axis for primary stars from $1.2$ (top) to $2.0\Msun$ (bottom) and energy transfer efficiency from $1.0$ (left) to $0.2$ (right). Three regimes are identified: no engulfment (blue), engulfment without envelope ejection (yellow; $\alpha E_\orb < |E_\bind|$), and successful envelope ejection (red; $\alpha E_\orb > |E_\bind|$). Crosses indicate results from 1D orbital integrations. The dashed line at $13\,\Mj$ marks the conventional boundary between planets and brown dwarfs.}
		\label{fig:engulf_regions_all_scenarios}
	\end{figure*}

As discussed in the previous section and illustrated in Fig.~\ref{fig:engulf_regions_all_scenarios}, engulfment near the RGB tip requires initial orbital separations in the range $\sim 0.5 - 3.5$ au, but the detection of low-mass companions around their main sequence host is observationally challenging. Transit and direct imaging surveys are poorly sensitive to such separations, while radial-velocity detections require long temporal baselines (orbital periods of $\sim 100$ days to several years). As a result, the occurrence of substellar companions in this parameter space remains poorly constrained.

\begin{figure}[h!]
		\centering
		\includegraphics[width=0.49\textwidth]{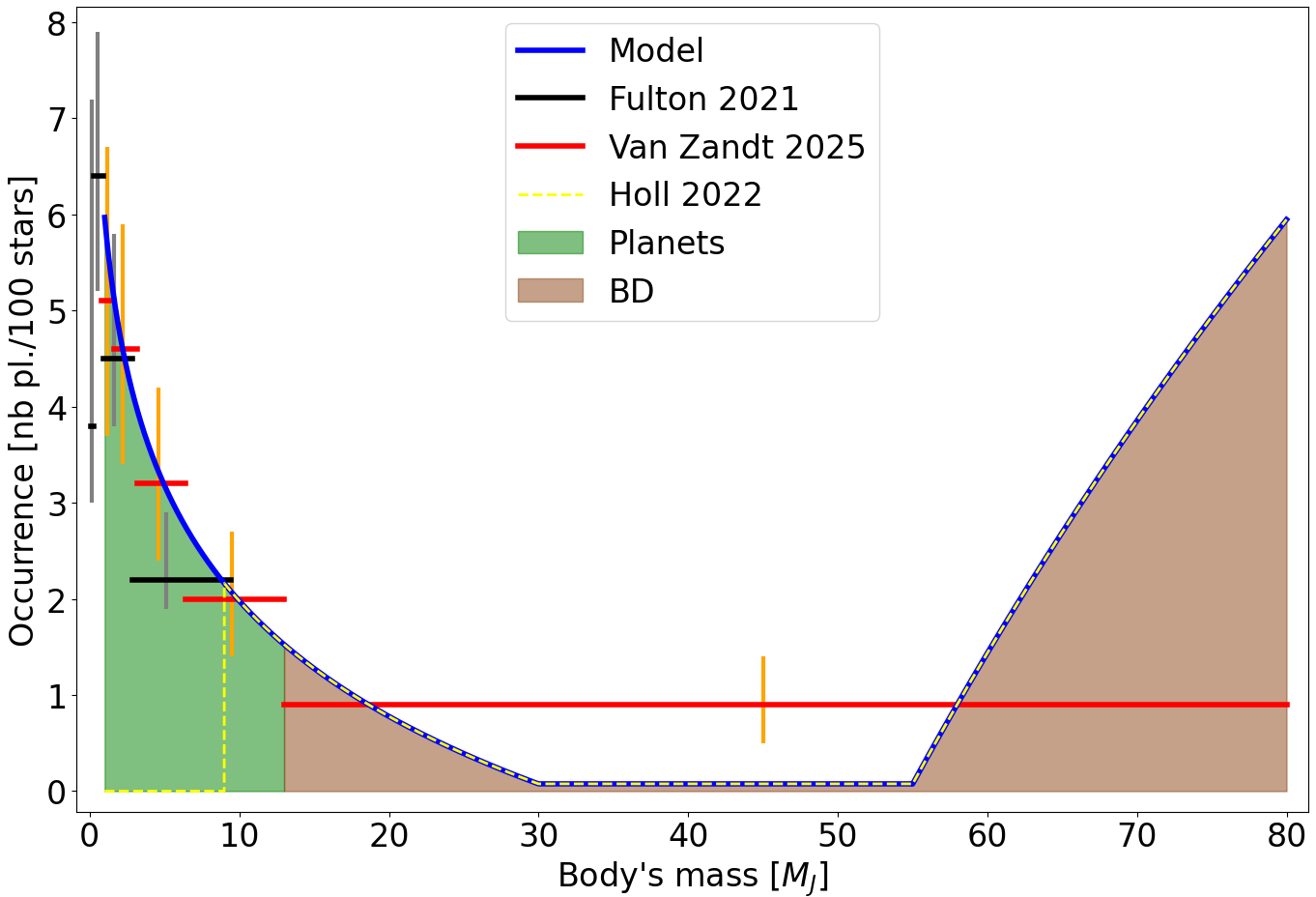}
		\caption{Mass distribution of substellar bodies around solar-type MS stars. Horizontal black and red lines show data from \cite{Fulton21} and \cite{Van_Zandt25}, with error bars in grey and orange, respectively. The yellow dashed line represents the distribution used by \cite{Holl22}, and the blue line is the model used in this work, extending Holl's distribution to planetary-mass objects. The green and brown zones represent mass ranges for planets and BDs, respectively.}
		\label{fig:eta M}
    \end{figure}

To model the underlying population of substellar companions, we need empirical distributions for the semi-major axis, mass, and host stellar mass. The semi-major axis distribution is adopted from \citet{Fulton21}, as refined by \citet{Van_Zandt25}
   \begin{equation}
		\eta(a) = 82 \times a^{-0.86} \times \left(1 - \exp\left[- \left( \frac{a}{5} \right)^{1.59}\right]\right) ~.
		\label{eq:distri a}
	\end{equation}
The companion mass distribution is based on \cite{Holl22} for brown dwarfs, extended to planetary masses using results from \citet{Fulton21} and \cite{Van_Zandt25} as shown in Fig.~\ref{fig:eta M} and is parametrized as
    \begin{eqnarray}
		\eta(M_{\ 1-30}) & = & -3.99 \times \log_{10}(M_\star) + 5.969 ~, 
		\label{eq:distri M_BD 10 30} \\
		\eta(M_{30-55}) & = &  0.076 ~,
		\label{eq:distri M_BD 30 55} \\
		\eta(M_{55-80}) & = &  36.1 \times \log_{10}(M_\star) - 62.753 ~.
		\label{eq:distri M_BD 55 80}
	\end{eqnarray}

The occurrence of substellar companions as a function of stellar mass is taken from \cite{Reffert15}, assuming solar metallicity,
   \begin{equation}
		\eta(M_\star) = 0.082 \times \exp \left[-\frac{1}{2} \left(\frac{M_\star - 1.8}{0.5}\right)^2 \right] ~.
		\label{eq:distri Ms}
	\end{equation}
This gives $\sim 4\,-\,8$ substellar companions per 100 stars in the $1.2-2.0\Msun$ range, with a peak around $M_\star \simeq 1.7\Msun$ (Table~\ref{tab:occurrence}).
    \begin{table}[h!]
		\begin{center}
		\begin{tabular}{|c|c|}
		
		\hline
		$M_\star [M_\odot]$ & nb bodies/100 stars \\
		\hline
		1.2   & $3.99$ \\
		1.5   & $6.85$ \\
		1.7   & $8.04$ \\
		2.0   & $7.57$ \\
		\hline
		\end{tabular}
		\end{center}
		\caption{Average number of substellar companions per 100 stars as a function of stellar mass, derived from Eq.~\ref{eq:distri Ms}.}
		\label{tab:occurrence}
	\end{table}

\begin{figure}[h!]
		\centering
		\includegraphics[width=0.49\textwidth]{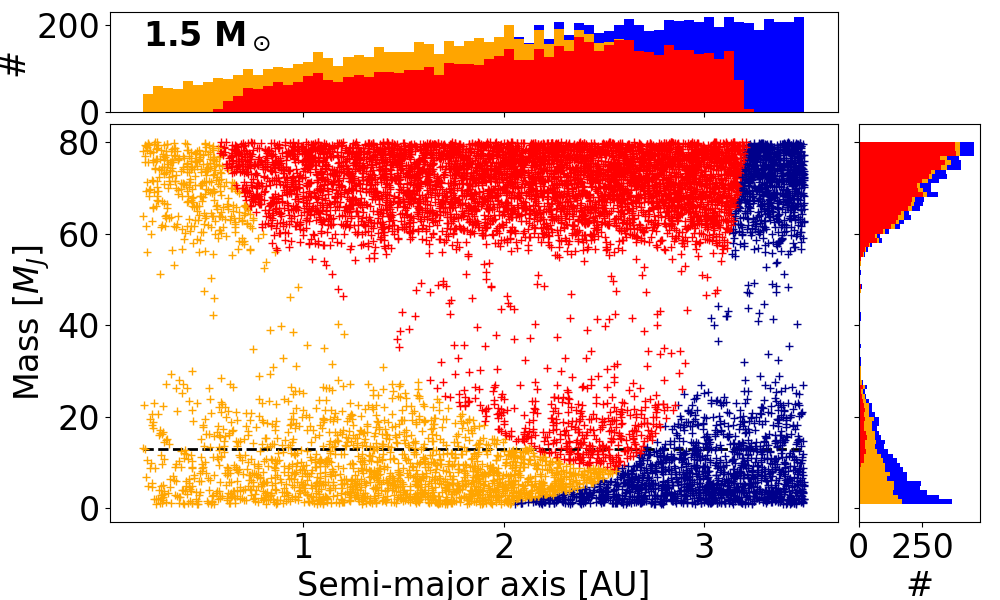}
		\caption{Fate of planets and BDs as a function of its mass and semi-major axis for a $1.5\Msun$ star and $\alpha = 1.0$. The top and right panels show the corresponding distributions. Blue crosses indicate bodies that avoid engulfment, orange crosses those engulfed but with insufficient energy deposition for envelope ejection, and red crosses those capable of expelling the envelope. The black dashed line marks the conventional boundary between planets and BD at 13 $\Mj$. Results are shown for a set of 10,000 configurations. Figure \ref{fig:generated allstar} shows similar results for all stellar masses and energy transfer efficiencies studied in this work.}
		\label{fig:generated Ms1.5}
	\end{figure}

For each stellar mass, we generate synthetic populations of $10^4$ companions by randomly sampling the $(M_{p}, a)$ parameter space according to these distributions (Eq.~\ref{eq:distri a} $-$ \ref{eq:distri M_BD 55 80}). The fate of each system is then determined using the criteria established in Sect.~\ref{sec: Energetical arguments}. The resulting distributions are illustrated in Fig.~\ref{fig:generated Ms1.5} for the $1.5 M_\odot$ and $\alpha=1$ case, and in Fig.~\ref{fig:generated allstar} for all values of $M_\star$ and $\alpha$. The number of systems in each outcome category is averaged over five realisations to reduce statistical fluctuations. These numbers are then weighted by the global occurrence of companions around stars of the corresponding mass (Eq.~\ref{eq:distri Ms}).

Assuming solar metallicity, Eq.~\ref{eq:distri Ms} indicates an average of $6.85$ substellar objects per 100 stars of $1.5\Msun$ (Eq.~\ref{eq:distri Ms}), thus, the 10,000 bodies we generated correspond to a sample of approximately 146,000 stars. According to the mass distribution (Eqs.~\ref{eq:distri M_BD 10 30}, \ref{eq:distri M_BD 30 55} and \ref{eq:distri M_BD 55 80}), this sample comprises 2694 planets and 7306 BDs on average. Of these, in the case $\alpha=1$, we determine that only 79 planets and 5409 BDs possess [$M_p, a$] parameters that place them within the envelope ejection zone. This implies that, within our sample, 5488 of the 146,000 $1.5\Msun$ stars will engulf a body sufficiently close to the RGB tip to expel their envelope, yielding an occurrence of approximately $3.74\,\%$. This contribution is overwhelmingly dominated by brown dwarfs ($\sim 3.69\,\%$), while planets account for only $\sim 0.05\,\%$. These values are for a uniform sample of 1.5 $M_\odot$ stars. For a global estimate of the occurrence of envelope-ejection events, the results obtained for each stellar mass must be weighted according to the stellar Initial Mass Function (IMF) (\citealp{Salpeter55}; see also \citealp{Bastian10}), which accounts for the relative abundance of stars formed at different initial masses. For the IMF weighting, each of our four stellar models is assumed to be representative of a finite progenitor-mass bin centred around the model mass, as listed in Table \ref{tab:Ms_weights}. The 1.2, 1.5, 1.7, and 2.0$\Msun$ models therefore represent progenitors over the corresponding mass intervals. We explicitly checked that the exact mass ranges for defining the bins only marginally influence the resulting occurrences. The weighting coefficients presented in Table \ref{tab:Ms_weights} are derived from the \cite{Salpeter55} IMF.

Table \ref{tab:global eject occurrence field star} summarizes the resulting IMF-weighted contributions of each stellar-mass bin to the occurrence of envelope-ejection events. Summing these contributions over all stellar-mass bins gives the global occurrence of envelope-ejection events, reported in Table \ref{tab:global_env_eject}, with separated contributions for BDs and planets. Our results indicate that $\sim 3.1\,\%$ of stars in the progenitor-mass bins represented by
our $1.2 - 2.0 \Msun$ models could eject their envelope through substellar companion engulfment, assuming an energy transfer efficiency $\alpha = 1$. This number drops to $\sim2.5\,\%$ and $\sim1.5\,\%$ for $\alpha = 0.5$ and $0.2$, respectively. As expected, the probability of envelope ejection decreases with increasing stellar mass, reflecting the larger binding energies of more massive RGB envelopes and the lower number of massive stars. Ejection probability is also reduced with decreasing energy-transfer efficiency, as a smaller fraction of the released orbital energy is available to unbind the envelope. For example, reducing the efficiency from $\alpha=1$ to $0.5$ lowers the fraction of systems undergoing envelope ejection from $3.1$ to $\sim 2.5\%$, almost entirely due to brown dwarf companions, with planets contributing only $\sim 0.02\%$. For $\alpha=0.2$, the fraction decreases further to $\sim 1.5\%$, with envelope ejection occurring exclusively in systems containing brown dwarf companions (Table \ref{tab:global_env_eject}).

\begin{table}[h!]
		\begin{center}
		\begin{tabular}{|c|c|c|}
		\hline
		Reference mass & Bin & Weighting coef. \\
        ($M_\odot$) & ($M_\odot$) & \\
		\hline
        1.2   & $1.05 - 1.35$ & $0.464$ \\
		1.5   & $1.35 - 1.60$ & $0.235$ \\
		1.7   & $1.60 - 1.85$ & $0.163$ \\
		2.0   & $1.85 - 2.15$ & $0.138$ \\
		\hline
		\end{tabular}
		\end{center}
		\caption{Weighting coefficients used for each stellar mass bin using the \cite{Salpeter55} IMF.}
		\label{tab:Ms_weights}
	\end{table}

\begin{table}[h!]
		\begin{center}
		\begin{tabular}{|c|c|c|c|c|}
		\hline
		$\alpha$ & \text{$M_\star [M_\odot]$} & \text{Planets} & \text{BD} & \text{Both}\\
        \hline
		1.0 & 1.2 & 0.055 & 1.165 & 1.220 \\
        1.0 & 1.5 & 0.012 & 0.870 & 0.882 \\
        1.0 & 1.7 & 0.003 & 0.633 & 0.636 \\
        1.0 & 2.0 & 0.000 & 0.357 & 0.357 \\
		\hline
        0.5 & 1.2 & 0.021 & 1.025 & 1.046 \\
        0.5 & 1.5 & 0.000 & 0.704 & 0.705 \\
        0.5 & 1.7 & 0.000 & 0.483 & 0.483 \\
        0.5 & 2.0 & 0.000 & 0.228 & 0.228 \\
		\hline
        0.2 & 1.2 & 0.002 & 0.756 & 0.758 \\
        0.2 & 1.5 & 0.000 & 0.436 & 0.436 \\
        0.2 & 1.7 & 0.000 & 0.258 & 0.258 \\
        0.2 & 2.0 & 0.000 & 0.037 & 0.037 \\
		\hline
		\end{tabular}
		\end{center}
		\caption{IMF-weighted contribution of each stellar mass to the occurrence of envelope-ejection events triggered by substellar companion engulfment, for different energy-transfer efficiencies.}
		\label{tab:global eject occurrence field star}
\end{table}

    \begin{table}[h!]
		\begin{center}
		\begin{tabular}{|c|c|c|c|}
		\hline
		$\alpha$ & BD+planets & BD & planets \\
		\hline
        1.0   & $3.094$ & $3.025$ & $0.069$ \\
		0.5   & $2.461$ & $2.440$ & $0.021$\\
		0.2   & $1.489$ & $1.487$ & $0.002$\\
		\hline
		\end{tabular}
		\end{center}
		\caption{Global envelope ejection occurrences triggered by substellar companion engulfments for stars
        in the progenitor-mass bins represented by the $1.2 - 2.0\Msun$ models with energy transfer efficiencies between 0.2 and 1.0. See Table \ref{tab:global eject occurrence field star} for details on the contribution from each stellar mass.}
		\label{tab:global_env_eject}
	\end{table}

The predicted envelope-ejection occurrence depends strongly on the adopted mass distribution of substellar companions (Fig.~\ref{fig:generated allstar}). Companions in the brown-dwarf desert make a negligible contribution because of their low occurrence. Moreover, the minimum companion mass required to eject the envelope increases with stellar mass and decreases with the energy-transfer efficiency $\alpha$ (Fig.~\ref{fig:engulf_regions_all_scenarios}). Consequently, low-mass brown dwarfs and planets contribute appreciably only for the lowest-mass progenitors or when high values of $\alpha$ are assumed. In all cases, companions less massive than $\sim55\,\Mj$ account for only a small fraction of the predicted envelope-ejection events.

\section{From envelope ejection to the formation of sdB stars}
\label{sdBformation}

Envelope ejection is a necessary, but not sufficient condition for the formation of an sdB star. Helium must also ignite in the core; in other words, the core must have reached the minimum mass required for helium ignition by the time the envelope is ejected. If helium ultimately fails to ignite after envelope ejection, the star instead evolves into a low-mass helium white dwarf (He-WD). 

The minimum core mass required for helium ignition depends strongly on the assumptions adopted in the evolutionary models and on their input physics. \cite{Arancibia-Rojas24} showed that progenitors with lower metallicity require slightly more massive cores for helium ignition, while the inclusion of overshooting decreases the progenitor mass below which helium ignition occurs under degenerate conditions. Their models further show that the helium-ignition mass depends sensitively on the mass-loss history. If the envelope is removed sufficiently early, the remnant leaves the RGB and evolves toward the WD cooling track before helium ignites. However, in some cases, helium can still ignite later through a delayed flash, after further core contraction provides the additional compression required for ignition \citep{Arancibia-Rojas24}. This allows helium ignition to occur at a lower core mass. In the other cases, the core remains below the ignition threshold and the remnant evolves into a He-WD. 

For our 1.2$-$2.0$\Msun$ \texttt{STAREVOL} models and their associated assumptions (see Sect.~2), He ignition always occurs through a He-flash close to the RGB tip, with only a slight decrease in core degeneracy for our most massive progenitor at 2.0$\Msun$. Figure \ref{fig:Post-ejection_mass} shows the core mass reached by our \texttt{STAREVOL} models at the time of envelope ejection, for ejection times spanning the final 50~Myr before the RGB tip. We see that the core mass at the RGB tip reaches $\sim 0.475\Msun$ for all progenitor masses, except for the 2.0$\Msun$ model, for which it reaches only $\sim 0.463\Msun$, as a direct consequence of its lower core degeneracy. Using 0.45$\,\Msun$ as an indicative helium-ignition threshold \citep{Arancibia-Rojas24}, we find that a small fraction of envelope ejections result in the formation of an sdB star. Most envelope-ejection events occur before the core has reached the mass required for helium ignition, leading to the formation of a He-WD. The resulting fraction of sdB versus He-WD formation nevertheless depends sensitively on several assumptions in the stellar evolution
calculations (including overshooting) and mass-loss history, as underlined above. A detailed determination of this threshold, and therefore of the fraction of envelope-ejection events that successfully lead to sdB formation, will be addressed in a forthcoming paper.

Approximately 2\% of low- and intermediate-mass stars were historically estimated to evolve into sdB stars based on the birthrate  derived by \citet{Heber86}. More recent observational constraints, however, favour substantially lower values. Using a volume-limited sample of hot subdwarfs within 500 pc, \citet{Dawson24} derived a mid-plane space density of $\rho_0 = (6.03 \pm 0.51)\times10^{-7}\mathrm{pc^{-3}}$. Assuming a representative sdB lifetime on the EHB\footnote{Note the EHB lifetime is model-dependent and varies significantly with sdB mass \citep{Arancibia-Rojas24}, as well as with the treatment of mixing at the convective-core boundary. The birthrate derived here should therefore be regarded as an order-of-magnitude estimate rather than a precisely determined value.} of $\sim 100-150\,\mathrm{Myr}$ \cite[][their Fig.~10]{Arancibia-Rojas24}, this corresponds to a birthrate density of $\sim 4-6\times10^{-15}\mathrm{pc^{-3}yr^{-1}}$. For comparison, the local white dwarf formation rate is estimated at $\sim 10^{-12}\mathrm{pc^{-3}yr^{-1}}$ \citep[e.g.][]{Liebert05}. Assuming that the local sdB and white dwarf formation rates trace the same stellar population, the ratio of the sdB and white dwarf formation rates suggests that only $\sim 0.4-0.6~\%$ of white-dwarf progenitors pass through an sdB phase. This value is consistent with recent studies indicating that sdBs are significantly rarer than predicted by binary population synthesis models \citep[][]{Rodrigez-Segovia25}.

These observational estimates cannot be directly compared with the envelope-ejection occurrences derived here, since only a fraction of envelope-ejection events ultimately lead to helium ignition. Furthermore, a prediction of the present-day sdB formation rate through substellar companion engulfment requires  accounting for the Galactic star-formation history and the corresponding evolutionary times. These additional steps will be addressed in Paper II.

% #############################################################################################################
\section{Summary and Conclusions}
\label{sec: Conclusion}

In this work, we investigated the conditions under which the engulfment of substellar companions (planets and brown dwarfs) can provide sufficient energy to eject the envelope of RGB stars. We systematically explored a broad range of stellar ($1.2 - 2.0\,\Msun$) and companion ($1 - 80\,\Mj$) masses and several ejection efficiencies. This comprehensive parameter study extends previous investigations, which have generally considered a limited set of stellar and companion configurations.

We modelled the orbital evolution of substellar companions during the subgiant and RGB phases, and estimated the orbital energy released during the inspiral down to the disruption radius, defined as the larger of the Roche and virial radii. We compared this energy to the binding energy of the stellar envelope. With energy transfer efficiencies of $0.2, 0.5$ and $1$, we identified the regions of parameter space where envelope ejection is energetically possible.

Our results show that successful envelope ejection is restricted to a relatively narrow region of parameter space, requiring (i) sufficiently massive companions and (ii) engulfment occurring very close to the RGB tip, where the envelope binding energy is minimal. Reducing the energy transfer efficiency further narrows this region. For a $1.5\Msun$ star with $\alpha = 1$, we find that companions more massive than $\sim 8\,\Mj$ can in principle unbind the envelope if engulfment occurs within the last $\sim 0.2\,\mathrm{Myr}$ before the RGB tip. This critical mass increases to $\sim 12\,\Mj$ for $\alpha=0.5$ and $\sim 24\,\Mj$ for $\alpha=0.2$. In general, envelope ejection is favoured in lower-mass stars because they reach larger radii and have less tightly bound envelopes, and for higher values of $\alpha$ and $M_p$.

These results are broadly consistent with previous studies, but highlight significant quantitative differences. While some studies show that only a fraction of the envelope is unbound \citep[e.g.][]{Staff16, OConnor23}, other studies suggested that the minimum companion mass required to eject the envelope ranges from $\sim 10\,\Mj$ \citep{Yarza23} to $\sim 30\,\Mj$ \citep{Kramer20} can expell the envelope. Our results reconcile part of this diversity by showing that the outcome depends sensitively on the evolutionary stage at engulfment and on the stellar mass, which were not always explored systematically in earlier works.

Several simplifying assumptions underlie our approach. Our treatment does not include hydrodynamical effects, angular momentum transport, or detailed mass-loss processes during the common-envelope phase. The disruption of the companion is treated in a simplified manner, assuming chemically homogeneous bodies, whereas real objects are likely to be differentiated and subject to complex ablation processes. Hydrodynamical simulations have shown that engulfed BD and planets can be ablated down to 10 \% of their initial mass before reaching their disruption radius \citep{Lau25, Lau26}, although technical limitations often prevent simulations from precisely locating $R_\disrupt$ due to disruption happening too deep in the star \citep{Kramer20, Staff16, Yarza23}. Analytical studies \citep{OConnor23, Nordhaus06} showed that for giant stars and massive companions the location of the disruption of BD and massive planets is their Roche radius. We also ignore additional energy sources, such as recombination energy, which could facilitate envelope ejection.

In addition, our estimate of the occurrence of envelope-ejection events relied on empirical distributions of substellar companion mass and orbital separation that remain uncertain, particularly in the brown dwarf desert and at separations of a few astronomical units. We also assumed that these distributions are independent, and that the orbital evolution is not affected by prior interactions or stellar multiplicity.

Combining our envelope-ejection criteria with the observed distributions of substellar companions, we estimate that engulfment could lead to envelope ejection in $\sim1.5-3.1\%$ of stars in the progenitor-mass bins represented by our $1.2-2.0\Msun$ models, depending on the assumed energy-transfer efficiency. Envelope ejection, however, does not necessarily lead to the formation of an sdB star: the core must also be sufficiently massive to ignite helium. The minimum core mass required for helium ignition depends strongly on the stellar mass, mass-loss history, and on the assumptions adopted in the stellar evolution calculations, including metallicity, overshooting, and other input physics. We postpone such calculations in a second paper, which will address the fraction of envelope-ejection events that successfully lead to sdB formation.

Overall, the present study highlights the critical importance of the timing of engulfment, companion mass, and energy transfer efficiency in determining the outcome of the interaction.  Further work, within the same framework and assumptions as adopted here, should explore the impact of additional parameters, such as stellar metallicity and orbital eccentricity. By mapping the regions of parameter space where envelope ejection is energetically possible, we provide a guide for future multidimensional hydrodynamical simulations, which are required to constrain the value of the energy transfer efficiency and to determine the actual fraction of systems that successfully eject their envelope.

% ##########################################################################################################
\begin{acknowledgements}

L.S. is a F.R.S.-FNRS Research Director. V.V.G. is a F.R.S.-FNRS Senior Research Associate. A.T. thanks the sdOB12 conference attendees for the interesting discussions, which encouraged us to explore more deeply several aspects of this topic. We also thank the anonymous referee for their contribution to the quality of this work. LLM were used in the writing of this document.

\end{acknowledgements}
 % ##########################################################################################################

\bibliography{References.bib}

@ARTICLE{Villaver14,
       author = {{Villaver}, Eva and {Livio}, Mario and {Mustill}, Alexander J. and {Siess}, Lionel},
        title = "{Hot Jupiters and Cool Stars}",
      journal = {\apj},
         year = 2014,
        month = oct,
       volume = {794},
       number = {1},
          eid = {3},
        pages = {3},
          doi = {10.1088/0004-637X/794/1/3},
archivePrefix = {arXiv},
       eprint = {1407.7879},
 primaryClass = {astro-ph.EP},
       adsurl = {https://ui.adsabs.harvard.edu/abs/2014ApJ...794....3V}
}

@ARTICLE{Villaver09,
       author = {{Villaver}, Eva and {Livio}, Mario},
        title = "{The Orbital Evolution of Gas Giant Planets Around Giant Stars}",
      journal = {\apjl},
         year = 2009,
        month = nov,
       volume = {705},
       number = {1},
        pages = {L81-L85},
          doi = {10.1088/0004-637X/705/1/L81},
archivePrefix = {arXiv},
       eprint = {0910.2396},
 primaryClass = {astro-ph.EP},
       adsurl = {https://ui.adsabs.harvard.edu/abs/2009ApJ...705L..81V}
}

@ARTICLE{Schroeder05,
       author = {{Schr{\"o}der}, K. -P. and {Cuntz}, M.},
        title = "{A New Version of Reimers' Law of Mass Loss Based on a Physical Approach}",
      journal = {\apjl},
         year = 2005,
        month = sep,
       volume = {630},
       number = {1},
        pages = {L73-L76},
          doi = {10.1086/491579},
archivePrefix = {arXiv},
       eprint = {astro-ph/0507598},
 primaryClass = {astro-ph},
       adsurl = {https://ui.adsabs.harvard.edu/abs/2005ApJ...630L..73S}
}

@ARTICLE{Metzger12,
       author = {{Metzger}, B.~D. and {Giannios}, D. and {Spiegel}, D.~S.},
        title = "{Optical and X-ray transients from planet-star mergers}",
      journal = {\mnras},
         year = 2012,
        month = oct,
       volume = {425},
       number = {4},
        pages = {2778-2798},
          doi = {10.1111/j.1365-2966.2012.21444.x},
archivePrefix = {arXiv},
       eprint = {1204.0796},
 primaryClass = {astro-ph.EP},
       adsurl = {https://ui.adsabs.harvard.edu/abs/2012MNRAS.425.2778M}
}

@ARTICLE{Staff16,
       author = {{Staff}, Jan E. and {De Marco}, Orsola and {Wood}, Peter and {Galaviz}, Pablo and {Passy}, Jean-Claude},
        title = "{Hydrodynamic simulations of the interaction between giant stars and planets}",
      journal = {\mnras},
         year = 2016,
        month = may,
       volume = {458},
       number = {1},
        pages = {832-844},
          doi = {10.1093/mnras/stw331},
archivePrefix = {arXiv},
       eprint = {1602.03130},
 primaryClass = {astro-ph.SR},
       adsurl = {https://ui.adsabs.harvard.edu/abs/2016MNRAS.458..832S}
}

@ARTICLE{OConnor23,
       author = {{O'Connor}, Christopher E. and {Bildsten}, Lars and {Cantiello}, Matteo and {Lai}, Dong},
        title = "{Giant Planet Engulfment by Evolved Giant Stars: Light Curves, Asteroseismology, and Survivability}",
      journal = {\apj},
         year = 2023,
        month = jun,
       volume = {950},
       number = {2},
          eid = {128},
        pages = {128},
          doi = {10.3847/1538-4357/acd2d4},
archivePrefix = {arXiv},
       eprint = {2304.09882},
 primaryClass = {astro-ph.EP},
       adsurl = {https://ui.adsabs.harvard.edu/abs/2023ApJ...950..128O}
}

@ARTICLE{Siess99a,
       author = {{Siess}, Lionel and {Livio}, Mario},
        title = "{The accretion of brown dwarfs and planets by giant stars - I. Asymptotic giant branch stars}",
      journal = {\mnras},
         year = 1999,
        month = apr,
       volume = {304},
       number = {4},
        pages = {925-937},
          doi = {10.1046/j.1365-8711.1999.02376.x},
archivePrefix = {arXiv},
       eprint = {astro-ph/9812220},
 primaryClass = {astro-ph},
       adsurl = {https://ui.adsabs.harvard.edu/abs/1999MNRAS.304..925S}
}

@ARTICLE{Siess99b,
       author = {{Siess}, Lionel and {Livio}, Mario},
        title = "{The accretion of brown dwarfs and planets by giant stars - II. Solar-mass stars on the red giant branch}",
      journal = {\mnras},
         year = 1999,
        month = oct,
       volume = {308},
       number = {4},
        pages = {1133-1149},
          doi = {10.1046/j.1365-8711.1999.02784.x},
archivePrefix = {arXiv},
       eprint = {astro-ph/9905235},
 primaryClass = {astro-ph},
       adsurl = {https://ui.adsabs.harvard.edu/abs/1999MNRAS.308.1133S}
}

@ARTICLE{Siess00,
       author = {{Siess}, L. and {Dufour}, E. and {Forestini}, M.},
        title = "{An internet server for pre-main sequence tracks of low- and intermediate-mass stars}",
      journal = {\aap},
         year = 2000,
        month = jun,
       volume = {358},
        pages = {593-599},
          doi = {10.48550/arXiv.astro-ph/0003477},
archivePrefix = {arXiv},
       eprint = {astro-ph/0003477},
 primaryClass = {astro-ph},
       adsurl = {https://ui.adsabs.harvard.edu/abs/2000A&A...358..593S}
}

@ARTICLE{Livio1988,
       author = {{Livio}, Mario and {Soker}, Noam},
        title = "{The Common Envelope Phase in the Evolution of Binary Stars}",
      journal = {\apj},
         year = 1988,
        month = jun,
       volume = {329},
        pages = {764},
          doi = {10.1086/166419},
       adsurl = {https://ui.adsabs.harvard.edu/abs/1988ApJ...329..764L}
}

@ARTICLE{Siess06,
       author = {{Siess}, L.},
        title = "{Evolution of massive AGB stars. I. Carbon burning phase}",
      journal = {\aap},
         year = 2006,
        month = mar,
       volume = {448},
       number = {2},
        pages = {717-729},
          doi = {10.1051/0004-6361:20053043},
       adsurl = {https://ui.adsabs.harvard.edu/abs/2006A&A...448..717S}
}

@ARTICLE{Ostriker99,
       author = {{Ostriker}, Eve C.},
        title = "{Dynamical Friction in a Gaseous Medium}",
      journal = {\apj},
         year = 1999,
        month = mar,
       volume = {513},
       number = {1},
        pages = {252-258},
          doi = {10.1086/306858},
archivePrefix = {arXiv},
       eprint = {astro-ph/9810324},
 primaryClass = {astro-ph},
       adsurl = {https://ui.adsabs.harvard.edu/abs/1999ApJ...513..252O}
}

@ARTICLE{Asplund09,
       author = {{Asplund}, Martin and {Grevesse}, Nicolas and {Sauval}, A. Jacques and {Scott}, Pat},
        title = "{The Chemical Composition of the Sun}",
      journal = {\araa},
         year = 2009,
        month = sep,
       volume = {47},
       number = {1},
        pages = {481-522},
          doi = {10.1146/annurev.astro.46.060407.145222},
archivePrefix = {arXiv},
       eprint = {0909.0948},
 primaryClass = {astro-ph.SR},
       adsurl = {https://ui.adsabs.harvard.edu/abs/2009ARA&A..47..481A}
}

@ARTICLE{Hut81,
       author = {{Hut}, P.},
        title = "{Tidal evolution in close binary systems.}",
      journal = {\aap},
         year = 1981,
        month = jun,
       volume = {99},
        pages = {126-140},
       adsurl = {https://ui.adsabs.harvard.edu/abs/1981A&A....99..126H}
}

@ARTICLE{Fellay23,
       author = {{Fellay}, L. and {Pezzotti}, C. and {Buldgen}, G. and {Eggenberger}, P. and {Bolmont}, E.},
        title = "{Constraints on planetary tidal dissipation from a detailed study of Kepler 91b}",
      journal = {\aap},
         year = 2023,
        month = jan,
       volume = {669},
          eid = {A2},
        pages = {A2},
          doi = {10.1051/0004-6361/202243621},
archivePrefix = {arXiv},
       eprint = {2210.11535},
 primaryClass = {astro-ph.EP},
       adsurl = {https://ui.adsabs.harvard.edu/abs/2023A&A...669A...2F}
}

@BOOK{Kepler1619,
       author = {{Kepler}, Johannes},
        title = "{Ioannis Keppleri harmonices mundi libri V : quorum primus harmonicus ... quartus metaphysicus, psychologicus et astrologicus geometricus ... secundus architectonicus ... tertius proprie ... quintus astronomicus \& metaphysicus ... : appendix habet comparationem huius operis cum harmonices Cl. Ptolemaei libro III cumque Roberti de Fluctibus ... speculationibus harmonicis, operi de macrocosmo \& microcosmo insertis}",
         year = 1619,
          doi = {10.3931/e-rara-8723},
       adsurl = {https://ui.adsabs.harvard.edu/abs/1619ikhm.book.....K}
}

@ARTICLE{Bolmont16,
       author = {{Bolmont}, Emeline and {Mathis}, St{\'e}phane},
        title = "{Effect of the rotation and tidal dissipation history of stars on the evolution of close-in planets}",
      journal = {Celestial Mechanics and Dynamical Astronomy},
         year = 2016,
        month = nov,
       volume = {126},
       number = {1-3},
        pages = {275-296},
          doi = {10.1007/s10569-016-9690-3},
archivePrefix = {arXiv},
       eprint = {1603.06268},
 primaryClass = {astro-ph.EP},
       adsurl = {https://ui.adsabs.harvard.edu/abs/2016CeMDA.126..275B}
}

@INPROCEEDINGS{Han13,
       author = {{Han}, Z. and {Chen}, X.},
        title = "{Envelope loss of RGB/AGB stars and the formation of hot subdwarfs}",
    booktitle = {European Physical Journal Web of Conferences},
         year = 2013,
       series = {European Physical Journal Web of Conferences},
       volume = {43},
        month = mar,
          eid = {01007},
        pages = {01007},
          doi = {10.1051/epjconf/20134301007},
       adsurl = {https://ui.adsabs.harvard.edu/abs/2013EPJWC..4301007H}
}

@ARTICLE{Fulton21,
       author = {{Fulton}, Benjamin J. and {Rosenthal}, Lee J. and {Hirsch}, Lea A. and {Isaacson}, Howard and {Howard}, Andrew W. and {Dedrick}, Cayla M. and {Sherstyuk}, Ilya A. and {Blunt}, Sarah C. and {Petigura}, Erik A. and {Knutson}, Heather A. and {Behmard}, Aida and {Chontos}, Ashley and {Crepp}, Justin R. and {Crossfield}, Ian J.~M. and {Dalba}, Paul A. and {Fischer}, Debra A. and {Henry}, Gregory W. and {Kane}, Stephen R. and {Kosiarek}, Molly and {Marcy}, Geoffrey W. and {Rubenzahl}, Ryan A. and {Weiss}, Lauren M. and {Wright}, Jason T.},
        title = "{California Legacy Survey. II. Occurrence of Giant Planets beyond the Ice Line}",
      journal = {\apjs},
         year = 2021,
        month = jul,
       volume = {255},
       number = {1},
          eid = {14},
        pages = {14},
          doi = {10.3847/1538-4365/abfcc1},
archivePrefix = {arXiv},
       eprint = {2105.11584},
 primaryClass = {astro-ph.EP},
       adsurl = {https://ui.adsabs.harvard.edu/abs/2021ApJS..255...14F}
}

@ARTICLE{Reffert15,
       author = {{Reffert}, Sabine and {Bergmann}, Christoph and {Quirrenbach}, Andreas and {Trifonov}, Trifon and {K{\"u}nstler}, Andreas},
        title = "{Precise radial velocities of giant stars. VII. Occurrence rate of giant extrasolar planets as a function of mass and metallicity}",
      journal = {\aap},
         year = 2015,
        month = feb,
       volume = {574},
          eid = {A116},
        pages = {A116},
          doi = {10.1051/0004-6361/201322360},
archivePrefix = {arXiv},
       eprint = {1412.4634},
 primaryClass = {astro-ph.EP},
       adsurl = {https://ui.adsabs.harvard.edu/abs/2015A&A...574A.116R}
}

@ARTICLE{Holl22,
       author = {{Holl}, B. and {Perryman}, M. and {Lindegren}, L. and {Segransan}, D. and {Raimbault}, M.},
        title = "{Gaia-predicted brown dwarf detection rates around FGK stars in astrometry, radial velocity, and photometric transits}",
      journal = {\aap},
         year = 2022,
        month = may,
       volume = {661},
          eid = {A151},
        pages = {A151},
          doi = {10.1051/0004-6361/201936393},
archivePrefix = {arXiv},
       eprint = {2109.02647},
 primaryClass = {astro-ph.SR},
       adsurl = {https://ui.adsabs.harvard.edu/abs/2022A&A...661A.151H}
}

@ARTICLE{Privitera16,
       author = {{Privitera}, Giovanni and {Meynet}, Georges and {Eggenberger}, Patrick and {Vidotto}, Aline A. and {Villaver}, Eva and {Bianda}, Michele},
        title = "{Star-planet interactions. I. Stellar rotation and planetary orbits}",
      journal = {\aap},
         year = 2016,
        month = jun,
       volume = {591},
          eid = {A45},
        pages = {A45},
          doi = {10.1051/0004-6361/201528044},
archivePrefix = {arXiv},
       eprint = {1604.06005},
 primaryClass = {astro-ph.EP},
       adsurl = {https://ui.adsabs.harvard.edu/abs/2016A&A...591A..45P}
}

@ARTICLE{Rasio96,
       author = {{Rasio}, F.~A. and {Tout}, C.~A. and {Lubow}, S.~H. and {Livio}, M.},
        title = "{Tidal Decay of Close Planetary Orbits}",
      journal = {\apj},
         year = 1996,
        month = oct,
       volume = {470},
        pages = {1187},
          doi = {10.1086/177941},
archivePrefix = {arXiv},
       eprint = {astro-ph/9605059},
 primaryClass = {astro-ph},
       adsurl = {https://ui.adsabs.harvard.edu/abs/1996ApJ...470.1187R}
}

@ARTICLE{Kramer20,
       author = {{Kramer}, M. and {Schneider}, F.~R.~N. and {Ohlmann}, S.~T. and {Geier}, S. and {Schaffenroth}, V. and {Pakmor}, R. and {R{\"o}pke}, F.~K.},
        title = "{Formation of sdB-stars via common envelope ejection by substellar companions}",
      journal = {\aap},
         year = 2020,
        month = oct,
       volume = {642},
          eid = {A97},
        pages = {A97},
          doi = {10.1051/0004-6361/202038702},
archivePrefix = {arXiv},
       eprint = {2007.00019},
 primaryClass = {astro-ph.SR},
       adsurl = {https://ui.adsabs.harvard.edu/abs/2020A&A...642A..97K}
}

@ARTICLE{Van_Grootel21,
       author = {{Van Grootel}, V. and {Pozuelos}, F.~J. and {Thuillier}, A. and {Charpinet}, S. and {Delrez}, L. and {Beck}, M. and {Fortier}, A. and {Hoyer}, S. and {Sousa}, S.~G. and {Barlow}, B.~N. and {Billot}, N. and {D{\'e}vora-Pajares}, M. and {{\O}stensen}, R.~H. and {Alibert}, Y. and {Alonso}, R. and {Anglada Escud{\'e}}, G. and {Asquier}, J. and {Barrado}, D. and {Barros}, S.~C.~C. and {Baumjohann}, W. and {Beck}, T. and {Bekkelien}, A. and {Benz}, W. and {Bonfils}, X. and {Brandeker}, A. and {Broeg}, C. and {Bruno}, G. and {B{\'a}rczy}, T. and {Cabrera}, J. and {Cameron}, A.~C. and {Charnoz}, S. and {Davies}, M.~B. and {Deleuil}, M. and {Demangeon}, O.~D.~S. and {Demory}, B. -O. and {Ehrenreich}, D. and {Erikson}, A. and {Fossati}, L. and {Fridlund}, M. and {Futyan}, D. and {Gandolfi}, D. and {Gillon}, M. and {Guedel}, M. and {Heng}, K. and {Isaak}, K.~G. and {Kiss}, L. and {Laskar}, J. and {Lecavelier des Etangs}, A. and {Lendl}, M. and {Lovis}, C. and {Magrin}, D. and {Maxted}, P.~F.~L. and {Mecina}, M. and {Mustill}, A.~J. and {Nascimbeni}, V. and {Olofsson}, G. and {Ottensamer}, R. and {Pagano}, I. and {Pall{\'e}}, E. and {Peter}, G. and {Piotto}, G. and {Plesseria}, J. -Y. and {Pollacco}, D. and {Queloz}, D. and {Ragazzoni}, R. and {Rando}, N. and {Rauer}, H. and {Ribas}, I. and {Santos}, N.~C. and {Scandariato}, G. and {S{\'e}gransan}, D. and {Silvotti}, R. and {Simon}, A.~E. and {Smith}, A.~M.~S. and {Steller}, M. and {Szab{\'o}}, G.~M. and {Thomas}, N. and {Udry}, S. and {Viotto}, V. and {Walton}, N.~A. and {Westerdorff}, K. and {Wilson}, T.~G.},
        title = "{A search for transiting planets around hot subdwarfs. I. Methods and performance tests on light curves from Kepler, K2, TESS, and CHEOPS}",
      journal = {\aap},
         year = 2021,
        month = jun,
       volume = {650},
          eid = {A205},
        pages = {A205},
          doi = {10.1051/0004-6361/202140381},
archivePrefix = {arXiv},
       eprint = {2104.10462},
 primaryClass = {astro-ph.EP},
       adsurl = {https://ui.adsabs.harvard.edu/abs/2021A&A...650A.205V}
}

@ARTICLE{Thuillier22,
       author = {{Thuillier}, A. and {Van Grootel}, V. and {D{\'e}vora-Pajares}, M. and {Pozuelos}, F.~J. and {Charpinet}, S. and {Siess}, L.},
        title = "{A search for transiting planets around hot subdwarfs. II. Supplementary methods and results from TESS Cycle 1}",
      journal = {\aap},
         year = 2022,
        month = aug,
       volume = {664},
          eid = {A113},
        pages = {A113},
          doi = {10.1051/0004-6361/202243554},
archivePrefix = {arXiv},
       eprint = {2209.02437},
 primaryClass = {astro-ph.EP},
       adsurl = {https://ui.adsabs.harvard.edu/abs/2022A&A...664A.113T}
}

@ARTICLE{Heber16,
       author = {{Heber}, U.},
        title = "{Hot Subluminous Stars}",
      journal = {\pasp},
         year = 2016,
        month = aug,
       volume = {128},
       number = {966},
        pages = {082001},
          doi = {10.1088/1538-3873/128/966/082001},
archivePrefix = {arXiv},
       eprint = {1604.07749},
 primaryClass = {astro-ph.SR},
       adsurl = {https://ui.adsabs.harvard.edu/abs/2016PASP..128h2001H}
}

@ARTICLE{Heber24,
       author = {{Heber}, Ulrich},
        title = "{Hot Subdwarf Stars}",
      journal = {arXiv e-prints},
         year = 2024,
        month = oct,
          eid = {arXiv:2410.11663},
        pages = {arXiv:2410.11663},
          doi = {10.48550/arXiv.2410.11663},
archivePrefix = {arXiv},
       eprint = {2410.11663},
 primaryClass = {astro-ph.SR},
       adsurl = {https://ui.adsabs.harvard.edu/abs/2024arXiv241011663H}
}

@ARTICLE{Iben90,
       author = {{Iben}, Jr., Icko},
        title = "{On the Consequences of Low-Mass White Dwarf Mergers}",
      journal = {\apj},
         year = 1990,
        month = apr,
       volume = {353},
        pages = {215},
          doi = {10.1086/168609},
       adsurl = {https://ui.adsabs.harvard.edu/abs/1990ApJ...353..215I}
}

@ARTICLE{Dorman93,
       author = {{Dorman}, Ben and {Rood}, Robert T. and {O'Connell}, Robert W.},
        title = "{Ultraviolet Radiation from Evolved Stellar Populations. I. Models}",
      journal = {\apj},
         year = 1993,
        month = dec,
       volume = {419},
        pages = {596},
          doi = {10.1086/173511},
archivePrefix = {arXiv},
       eprint = {astro-ph/9311022},
 primaryClass = {astro-ph},
       adsurl = {https://ui.adsabs.harvard.edu/abs/1993ApJ...419..596D}
}

@ARTICLE{Schaffenroth14,
       author = {{Schaffenroth}, V. and {Geier}, S. and {Heber}, U. and {Kupfer}, T. and {Ziegerer}, E. and {Heuser}, C. and {Classen}, L. and {Cordes}, O.},
        title = "{Binaries discovered by the MUCHFUSS project. SDSS J162256.66+473051.1: An eclipsing subdwarf B binary with a brown dwarf companion}",
      journal = {\aap},
         year = 2014,
        month = apr,
       volume = {564},
          eid = {A98},
        pages = {A98},
          doi = {10.1051/0004-6361/201423377},
archivePrefix = {arXiv},
       eprint = {1402.5263},
 primaryClass = {astro-ph.SR},
       adsurl = {https://ui.adsabs.harvard.edu/abs/2014A&A...564A..98S}
}

@ARTICLE{Schaffenroth15,
       author = {{Schaffenroth}, V. and {Barlow}, B.~N. and {Drechsel}, H. and {Dunlap}, B.~H.},
        title = "{An eclipsing post common-envelope system consisting of a pulsating hot subdwarf B star and a brown dwarf companion}",
      journal = {\aap},
         year = 2015,
        month = apr,
       volume = {576},
          eid = {A123},
        pages = {A123},
          doi = {10.1051/0004-6361/201525701},
archivePrefix = {arXiv},
       eprint = {1502.04459},
 primaryClass = {astro-ph.SR},
       adsurl = {https://ui.adsabs.harvard.edu/abs/2015A&A...576A.123S}
}

@ARTICLE{Schaffenroth18,
       author = {{Schaffenroth}, V. and {Geier}, S. and {Heber}, U. and {Gerber}, R. and {Schneider}, D. and {Ziegerer}, E. and {Cordes}, O.},
        title = "{The MUCHFUSS photometric campaign}",
      journal = {\aap},
         year = 2018,
        month = jun,
       volume = {614},
          eid = {A77},
        pages = {A77},
          doi = {10.1051/0004-6361/201629789},
archivePrefix = {arXiv},
       eprint = {1802.03260},
 primaryClass = {astro-ph.SR},
       adsurl = {https://ui.adsabs.harvard.edu/abs/2018A&A...614A..77S}
}

@ARTICLE{Schaffenroth21,
       author = {{Schaffenroth}, V. and {Casewell}, S.~L. and {Schneider}, D. and {Kilkenny}, D. and {Geier}, S. and {Heber}, U. and {Irrgang}, A. and {Przybilla}, N. and {Marsh}, T.~R. and {Littlefair}, S.~P. and {Dhillon}, V.~S.},
        title = "{A quantitative in-depth analysis of the prototype sdB+BD system SDSS J08205+0008 revisited in the Gaia era}",
      journal = {\mnras},
         year = 2021,
        month = mar,
       volume = {501},
       number = {3},
        pages = {3847-3870},
          doi = {10.1093/mnras/staa3661},
archivePrefix = {arXiv},
       eprint = {2011.10013},
 primaryClass = {astro-ph.SR},
       adsurl = {https://ui.adsabs.harvard.edu/abs/2021MNRAS.501.3847S}
}

@ARTICLE{Nordhaus06,
       author = {{Nordhaus}, J. and {Blackman}, E.~G.},
        title = "{Low-mass binary-induced outflows from asymptotic giant branch stars}",
      journal = {\mnras},
         year = 2006,
        month = aug,
       volume = {370},
       number = {4},
        pages = {2004-2012},
          doi = {10.1111/j.1365-2966.2006.10625.x},
archivePrefix = {arXiv},
       eprint = {astro-ph/0604445},
 primaryClass = {astro-ph},
       adsurl = {https://ui.adsabs.harvard.edu/abs/2006MNRAS.370.2004N}
}

@ARTICLE{Maxted01,
       author = {{Maxted}, P.~F.~L. and {Heber}, U. and {Marsh}, T.~R. and {North}, R.~C.},
        title = "{The binary fraction of extreme horizontal branch stars}",
      journal = {\mnras},
         year = 2001,
        month = oct,
       volume = {326},
       number = {4},
        pages = {1391-1402},
          doi = {10.1111/j.1365-2966.2001.04714.x},
archivePrefix = {arXiv},
       eprint = {astro-ph/0103342},
 primaryClass = {astro-ph},
       adsurl = {https://ui.adsabs.harvard.edu/abs/2001MNRAS.326.1391M}
}

@ARTICLE{Sweigart97,
       author = {{Sweigart}, Allen V.},
        title = "{Effects of Helium Mixing on the Evolution of Globular Cluster Stars}",
      journal = {\apjl},
         year = 1997,
        month = jan,
       volume = {474},
       number = {1},
        pages = {L23-L26},
          doi = {10.1086/310414},
       adsurl = {https://ui.adsabs.harvard.edu/abs/1997ApJ...474L..23S}
}

@ARTICLE{Schaffenroth22,
       author = {{Schaffenroth}, V. and {Pelisoli}, I. and {Barlow}, B.~N. and {Geier}, S. and {Kupfer}, T.},
        title = "{Hot subdwarfs in close binaries observed from space. I. Orbital, atmospheric, and absolute parameters, and the nature of their companions}",
      journal = {\aap},
         year = 2022,
        month = oct,
       volume = {666},
          eid = {A182},
        pages = {A182},
          doi = {10.1051/0004-6361/202244214},
archivePrefix = {arXiv},
       eprint = {2207.02001},
 primaryClass = {astro-ph.SR},
       adsurl = {https://ui.adsabs.harvard.edu/abs/2022A&A...666A.182S}
}

@ARTICLE{Geier13b,
       author = {{Geier}, S. and {Heber}, U. and {Heuser}, C. and {Classen}, L. and {O'Toole}, S.~J. and {Edelmann}, H.},
        title = "{The subdwarf B star SB 290 - A fast rotator on the extreme horizontal branch}",
      journal = {\aap},
         year = 2013,
        month = mar,
       volume = {551},
          eid = {L4},
        pages = {L4},
          doi = {10.1051/0004-6361/201220964},
archivePrefix = {arXiv},
       eprint = {1301.4129},
 primaryClass = {astro-ph.SR},
       adsurl = {https://ui.adsabs.harvard.edu/abs/2013A&A...551L...4G}
}

@ARTICLE{Geier11a,
       author = {{Geier}, S. and {Classen}, L. and {Heber}, U.},
        title = "{The Fast-rotating, Low-gravity Subdwarf B Star EC 22081-1916: Remnant of a Common Envelope Merger Event}",
      journal = {\apjl},
         year = 2011,
        month = may,
       volume = {733},
       number = {1},
          eid = {L13},
        pages = {L13},
          doi = {10.1088/2041-8205/733/1/L13},
archivePrefix = {arXiv},
       eprint = {1104.4202},
 primaryClass = {astro-ph.SR},
       adsurl = {https://ui.adsabs.harvard.edu/abs/2011ApJ...733L..13G}
}

@ARTICLE{Zhang12,
       author = {{Zhang}, Xianfei and {Jeffery}, C. Simon},
        title = "{Evolutionary models for double helium white dwarf mergers and the formation of helium-rich hot subdwarfs}",
      journal = {\mnras},
         year = 2012,
        month = jan,
       volume = {419},
       number = {1},
        pages = {452-464},
          doi = {10.1111/j.1365-2966.2011.19711.x},
       adsurl = {https://ui.adsabs.harvard.edu/abs/2012MNRAS.419..452Z}
}

@ARTICLE{Yu21,
       author = {{Yu}, Jinlong and {Zhang}, Xianfei and {L{\"u}}, Guoliang},
        title = "{Post-merger evolution of double helium white dwarfs and distribution of helium-rich hot subdwarfs}",
      journal = {\mnras},
         year = 2021,
        month = jun,
       volume = {504},
       number = {2},
        pages = {2670-2674},
          doi = {10.1093/mnras/stab1063},
archivePrefix = {arXiv},
       eprint = {2104.06817},
 primaryClass = {astro-ph.SR},
       adsurl = {https://ui.adsabs.harvard.edu/abs/2021MNRAS.504.2670Y}
}

@ARTICLE{Nelemans98,
       author = {{Nelemans}, G. and {Tauris}, T.~M.},
        title = "{Formation of undermassive single white dwarfs and the influence of planets on late stellar evolution}",
      journal = {\aap},
         year = 1998,
        month = jul,
       volume = {335},
        pages = {L85-L88},
          doi = {10.48550/arXiv.astro-ph/9806011},
archivePrefix = {arXiv},
       eprint = {astro-ph/9806011},
 primaryClass = {astro-ph},
       adsurl = {https://ui.adsabs.harvard.edu/abs/1998A&A...335L..85N}
}

@ARTICLE{Yarza23,
       author = {{Yarza}, Ricardo and {Razo-L{\'o}pez}, Naela B. and {Murguia-Berthier}, Ariadna and {Everson}, Rosa Wallace and {Antoni}, Andrea and {MacLeod}, Morgan and {Soares-Furtado}, Melinda and {Lee}, Dongwook and {Ramirez-Ruiz}, Enrico},
        title = "{Hydrodynamics and Survivability during Post-main-sequence Planetary Engulfment}",
      journal = {\apj},
         year = 2023,
        month = sep,
       volume = {954},
       number = {2},
          eid = {176},
        pages = {176},
          doi = {10.3847/1538-4357/acbdfc},
archivePrefix = {arXiv},
       eprint = {2203.11227},
 primaryClass = {astro-ph.SR},
       adsurl = {https://ui.adsabs.harvard.edu/abs/2023ApJ...954..176Y}
}

@ARTICLE{Copperwheat11,
       author = {{Copperwheat}, C.~M. and {Morales-Rueda}, L. and {Marsh}, T.~R. and {Maxted}, P.~F.~L. and {Heber}, U.},
        title = "{Radial-velocity measurements of subdwarf B stars}",
      journal = {\mnras},
         year = 2011,
        month = aug,
       volume = {415},
       number = {2},
        pages = {1381-1395},
          doi = {10.1111/j.1365-2966.2011.18786.x},
archivePrefix = {arXiv},
       eprint = {1103.4745},
 primaryClass = {astro-ph.SR},
       adsurl = {https://ui.adsabs.harvard.edu/abs/2011MNRAS.415.1381C}
}

@ARTICLE{Vos18,
       author = {{Vos}, Joris and {N{\'e}meth}, P{\'e}ter and {Vu{\v{c}}kovi{\'c}}, Maja and {{\O}stensen}, Roy and {Parsons}, Steven},
        title = "{Composite hot subdwarf binaries - I. The spectroscopically confirmed sdB sample}",
      journal = {\mnras},
         year = 2018,
        month = jan,
       volume = {473},
       number = {1},
        pages = {693-709},
          doi = {10.1093/mnras/stx2198},
archivePrefix = {arXiv},
       eprint = {1708.07340},
 primaryClass = {astro-ph.SR},
       adsurl = {https://ui.adsabs.harvard.edu/abs/2018MNRAS.473..693V}
}

@ARTICLE{Han02,
       author = {{Han}, Z. and {Podsiadlowski}, Ph. and {Maxted}, P.~F.~L. and {Marsh}, T.~R. and {Ivanova}, N.},
        title = "{The origin of subdwarf B stars - I. The formation channels}",
      journal = {\mnras},
         year = 2002,
        month = oct,
       volume = {336},
       number = {2},
        pages = {449-466},
          doi = {10.1046/j.1365-8711.2002.05752.x},
archivePrefix = {arXiv},
       eprint = {astro-ph/0206130},
 primaryClass = {astro-ph},
       adsurl = {https://ui.adsabs.harvard.edu/abs/2002MNRAS.336..449H}
}

@ARTICLE{Han03,
       author = {{Han}, Z. and {Podsiadlowski}, Ph. and {Maxted}, P.~F.~L. and {Marsh}, T.~R.},
        title = "{The origin of subdwarf B stars - II}",
      journal = {\mnras},
         year = 2003,
        month = may,
       volume = {341},
       number = {2},
        pages = {669-691},
          doi = {10.1046/j.1365-8711.2003.06451.x},
archivePrefix = {arXiv},
       eprint = {astro-ph/0301380},
 primaryClass = {astro-ph},
       adsurl = {https://ui.adsabs.harvard.edu/abs/2003MNRAS.341..669H}
}

@ARTICLE{Siess13,
       author = {{Siess}, L. and {Izzard}, R.~G. and {Davis}, P.~J. and {Deschamps}, R.},
        title = "{BINSTAR: a new binary stellar evolution code. Tidal interactions}",
      journal = {\aap},
         year = 2013,
        month = feb,
       volume = {550},
          eid = {A100},
        pages = {A100},
          doi = {10.1051/0004-6361/201220327},
       adsurl = {https://ui.adsabs.harvard.edu/abs/2013A&A...550A.100S}
}

@ARTICLE{Heber86,
       author = {{Heber}, U.},
        title = "{The atmosphere of subluminous B stars. II. Analysis of 10 helium poor subdwarfs and the birthrate of sdB stars.}",
      journal = {\aap},
         year = 1986,
        month = jan,
       volume = {155},
        pages = {33-45},
       adsurl = {https://ui.adsabs.harvard.edu/abs/1986A&A...155...33H}
}

@ARTICLE{Dawson24,
       author = {{Dawson}, H. and {Geier}, S. and {Heber}, U. and {Pelisoli}, I. and {Dorsch}, M. and {Schaffenroth}, V. and {Reindl}, N. and {Culpan}, R. and {Pritzkuleit}, M. and {Vos}, J. and {Soemitro}, A.~A. and {Roth}, M.~M. and {Schneider}, D. and {Uzundag}, M. and {Vu{\v{c}}kovi{\'c}}, M. and {Antunes Amaral}, L. and {Istrate}, A.~G. and {Justham}, S. and {{\O}stensen}, R.~H. and {Telting}, J.~H. and {Djupvik}, A.~A. and {Raddi}, R. and {Green}, E.~M. and {Jeffery}, C.~S. and {Kepler}, S.~O. and {Munday}, J. and {Steinmetz}, T. and {Kupfer}, T.},
        title = "{A 500 pc volume-limited sample of hot subluminous stars. I. Space density, scale height, and population properties}",
      journal = {\aap},
         year = 2024,
        month = jun,
       volume = {686},
          eid = {A25},
        pages = {A25},
          doi = {10.1051/0004-6361/202348319},
archivePrefix = {arXiv},
       eprint = {2403.15513},
 primaryClass = {astro-ph.SR},
       adsurl = {https://ui.adsabs.harvard.edu/abs/2024A&A...686A..25D}
}

@ARTICLE{Liebert05,
       author = {{Liebert}, James and {Bergeron}, P. and {Holberg}, J.~B.},
        title = "{The Formation Rate and Mass and Luminosity Functions of DA White Dwarfs from the Palomar Green Survey}",
      journal = {\apjs},
         year = 2005,
        month = jan,
       volume = {156},
       number = {1},
        pages = {47-68},
          doi = {10.1086/425738},
archivePrefix = {arXiv},
       eprint = {astro-ph/0406657},
 primaryClass = {astro-ph},
       adsurl = {https://ui.adsabs.harvard.edu/abs/2005ApJS..156...47L}
}

@ARTICLE{Heber09,
       author = {{Heber}, Ulrich},
        title = "{Hot Subdwarf Stars}",
      journal = {\araa},
         year = 2009,
        month = sep,
       volume = {47},
       number = {1},
        pages = {211-251},
          doi = {10.1146/annurev-astro-082708-101836},
       adsurl = {https://ui.adsabs.harvard.edu/abs/2009ARA&A..47..211H}
}

@ARTICLE{Soker98,
       author = {{Soker}, Noam},
        title = "{Can Planets Influence the Horizontal Branch Morphology?}",
      journal = {\aj},
         year = 1998,
        month = sep,
       volume = {116},
       number = {3},
        pages = {1308-1313},
          doi = {10.1086/300503},
archivePrefix = {arXiv},
       eprint = {astro-ph/9803223},
 primaryClass = {astro-ph},
       adsurl = {https://ui.adsabs.harvard.edu/abs/1998AJ....116.1308S}
}

@ARTICLE{Van_Zandt25,
       author = {{Van Zandt}, Judah and {Gilbert}, Greg and {Giacalone}, Steven and {Petigura}, Erik and {Howard}, Andrew and {Handley}, Luke},
        title = "{A Smooth Transition from Giant Planets to Brown Dwarfs from the Radial Occurrence Distribution}",
      journal = {arXiv e-prints},
         year = 2025,
        month = nov,
          eid = {arXiv:2511.18758},
        pages = {arXiv:2511.18758},
          doi = {10.48550/arXiv.2511.18758},
archivePrefix = {arXiv},
       eprint = {2511.18758},
 primaryClass = {astro-ph.EP},
       adsurl = {https://ui.adsabs.harvard.edu/abs/2025arXiv251118758V}
}

@ARTICLE{Bastian10,
       author = {{Bastian}, Nate and {Covey}, Kevin R. and {Meyer}, Michael R.},
        title = "{A Universal Stellar Initial Mass Function? A Critical Look at Variations}",
      journal = {\araa},
         year = 2010,
        month = sep,
       volume = {48},
        pages = {339-389},
          doi = {10.1146/annurev-astro-082708-101642},
archivePrefix = {arXiv},
       eprint = {1001.2965},
 primaryClass = {astro-ph.GA},
       adsurl = {https://ui.adsabs.harvard.edu/abs/2010ARA&A..48..339B}
}

@ARTICLE{Salpeter55,
       author = {{Salpeter}, Edwin E.},
        title = "{The Luminosity Function and Stellar Evolution.}",
      journal = {\apj},
         year = 1955,
        month = jan,
       volume = {121},
        pages = {161},
          doi = {10.1086/145971},
       adsurl = {https://ui.adsabs.harvard.edu/abs/1955ApJ...121..161S}
}

@ARTICLE{Fontaine12,
       author = {{Fontaine}, G. and {Brassard}, P. and {Charpinet}, S. and {Green}, E.~M. and {Randall}, S.~K. and {Van Grootel}, V.},
        title = "{A preliminary look at the empirical mass distribution of hot B subdwarf stars}",
      journal = {\aap},
         year = 2012,
        month = mar,
       volume = {539},
          eid = {A12},
        pages = {A12},
          doi = {10.1051/0004-6361/201118220},
       adsurl = {https://ui.adsabs.harvard.edu/abs/2012A&A...539A..12F}
}

@ARTICLE{Ratzloff19,
       author = {{Ratzloff}, Jeffrey K. and {Barlow}, Brad N. and {Kupfer}, Thomas and {Corcoran}, Kyle A. and {Geier}, Stephan and {Bauer}, Evan and {Corbett}, Henry T. and {Howard}, Ward S. and {Glazier}, Amy and {Law}, Nicholas M.},
        title = "{EVR-CB-001: An Evolving, Progenitor, White Dwarf Compact Binary Discovered with the Evryscope}",
      journal = {\apj},
         year = 2019,
        month = sep,
       volume = {883},
       number = {1},
          eid = {51},
        pages = {51},
          doi = {10.3847/1538-4357/ab3727},
archivePrefix = {arXiv},
       eprint = {1909.02012},
 primaryClass = {astro-ph.SR},
       adsurl = {https://ui.adsabs.harvard.edu/abs/2019ApJ...883...51R}
}

@ARTICLE{Latour26,
       author = {{Latour}, M. and {Green}, E.~M. and {Dorsch}, M. and {Van Grootel}, V. and {Chayer}, P. and {Charpinet}, S. and {Heber}, U. and {Randall}, S.~K. and {Ma}, X.-Y.},
        title = "{The Arizona-Montr{\'e}al Spectroscopic Survey of hot subluminous stars}",
      journal = {\aap},
         year = 2026,
        month = jan,
       volume = {705},
          eid = {A248},
        pages = {A248},
          doi = {10.1051/0004-6361/202557314},
archivePrefix = {arXiv},
       eprint = {2511.02539},
 primaryClass = {astro-ph.SR},
       adsurl = {https://ui.adsabs.harvard.edu/abs/2026A&A...705A.248L}
}

@ARTICLE{Geier12,
       author = {{Geier}, S. and {Heber}, U.},
        title = "{Hot subdwarf stars in close-up view. II. Rotational properties of single and wide binary subdwarf B stars}",
      journal = {\aap},
         year = 2012,
        month = jul,
       volume = {543},
          eid = {A149},
        pages = {A149},
          doi = {10.1051/0004-6361/201219463},
archivePrefix = {arXiv},
       eprint = {1206.6977},
 primaryClass = {astro-ph.SR},
       adsurl = {https://ui.adsabs.harvard.edu/abs/2012A&A...543A.149G}
}

@ARTICLE{Charpinet18,
       author = {{Charpinet}, Stephane and {Giammichele}, Noemi and {Zong}, Weikai and {Van Grootel}, Val{\'e}rie and {Brassard}, Pierre and {Fontaine}, Gilles},
        title = "{Rotation in sdB stars as revealed by stellar oscillations}",
      journal = {Open Astronomy},
         year = 2018,
        month = jul,
       volume = {27},
       number = {1},
        pages = {112-119},
          doi = {10.1515/astro-2018-0012},
       adsurl = {https://ui.adsabs.harvard.edu/abs/2018OAst...27..112C}
}

@ARTICLE{Mosser12,
       author = {{Mosser}, B. and {Goupil}, M.~J. and {Belkacem}, K. and {Marques}, J.~P. and {Beck}, P.~G. and {Bloemen}, S. and {De Ridder}, J. and {Barban}, C. and {Deheuvels}, S. and {Elsworth}, Y. and {Hekker}, S. and {Kallinger}, T. and {Ouazzani}, R.~M. and {Pinsonneault}, M. and {Samadi}, R. and {Stello}, D. and {Garc{\'\i}a}, R.~A. and {Klaus}, T.~C. and {Li}, J. and {Mathur}, S. and {Morris}, R.~L.},
        title = "{Spin down of the core rotation in red giants}",
      journal = {\aap},
         year = 2012,
        month = dec,
       volume = {548},
          eid = {A10},
        pages = {A10},
          doi = {10.1051/0004-6361/201220106},
archivePrefix = {arXiv},
       eprint = {1209.3336},
 primaryClass = {astro-ph.SR},
       adsurl = {https://ui.adsabs.harvard.edu/abs/2012A&A...548A..10M}
}

@ARTICLE{Lau25,
       author = {{Lau}, Mike Y.~M. and {Cantiello}, Matteo and {Jermyn}, Adam S. and {MacLeod}, Morgan and {Mandel}, Ilya and {Price}, Daniel J.},
        title = "{Hot Jupiter engulfment by an early red giant in 3D hydrodynamics}",
      journal = {\aap},
         year = 2025,
        month = feb,
       volume = {694},
          eid = {A264},
        pages = {A264},
          doi = {10.1051/0004-6361/202452081},
archivePrefix = {arXiv},
       eprint = {2210.15848},
 primaryClass = {astro-ph.SR},
       adsurl = {https://ui.adsabs.harvard.edu/abs/2025A&A...694A.264L}
}

@ARTICLE{Lau26,
       author = {{Lau}, Mike Y.~M. and {Andrassy}, Robert and {Leidi}, Giovanni and {Gagnier}, Damien and {Mor{\'a}n-Fraile}, Javier and {R{\"o}pke}, Friedrich K. and {Mandel}, Ilya},
        title = "{Continuous mass ablation of planets engulfed in stellar envelopes}",
      journal = {arXiv e-prints},
         year = 2026,
        month = may,
          eid = {arXiv:2605.11343},
        pages = {arXiv:2605.11343},
          doi = {10.48550/arXiv.2605.11343},
archivePrefix = {arXiv},
       eprint = {2605.11343},
 primaryClass = {astro-ph.SR},
       adsurl = {https://ui.adsabs.harvard.edu/abs/2026arXiv260511343L}
}

@ARTICLE{Jia18,
       author = {{Jia}, Shi and {Spruit}, H.~C.},
        title = "{Disruption of a Planet Spiraling into its Host Star}",
      journal = {\apj},
         year = 2018,
        month = sep,
       volume = {864},
       number = {2},
          eid = {169},
        pages = {169},
          doi = {10.3847/1538-4357/aad77c},
archivePrefix = {arXiv},
       eprint = {1808.00467},
 primaryClass = {astro-ph.EP},
       adsurl = {https://ui.adsabs.harvard.edu/abs/2018ApJ...864..169J}
}

@ARTICLE{Jones16,
       author = {{Jones}, M.~I. and {Jenkins}, J.~S. and {Brahm}, R. and {Wittenmyer}, R.~A. and {Olivares E.}, F. and {Melo}, C.~H.~F. and {Rojo}, P. and {Jord{\'a}n}, A. and {Drass}, H. and {Butler}, R.~P. and {Wang}, L.},
        title = "{Four new planets around giant stars and the mass-metallicity correlation of planet-hosting stars}",
      journal = {\aap},
         year = 2016,
        month = may,
       volume = {590},
          eid = {A38},
        pages = {A38},
          doi = {10.1051/0004-6361/201628067},
archivePrefix = {arXiv},
       eprint = {1603.03738},
 primaryClass = {astro-ph.EP},
       adsurl = {https://ui.adsabs.harvard.edu/abs/2016A&A...590A..38J}
}

@ARTICLE{Fischer05,
       author = {{Fischer}, Debra A. and {Valenti}, Jeff},
        title = "{The Planet-Metallicity Correlation}",
      journal = {\apj},
         year = 2005,
        month = apr,
       volume = {622},
       number = {2},
        pages = {1102-1117},
          doi = {10.1086/428383},
       adsurl = {https://ui.adsabs.harvard.edu/abs/2005ApJ...622.1102F}
}

@ARTICLE{Santos01,
       author = {{Santos}, N.~C. and {Israelian}, G. and {Mayor}, M.},
        title = "{The metal-rich nature of stars with planets}",
      journal = {\aap},
         year = 2001,
        month = jul,
       volume = {373},
        pages = {1019-1031},
          doi = {10.1051/0004-6361:20010648},
archivePrefix = {arXiv},
       eprint = {astro-ph/0105216},
 primaryClass = {astro-ph},
       adsurl = {https://ui.adsabs.harvard.edu/abs/2001A&A...373.1019S}
}

@ARTICLE{Gonzalez97,
       author = {{Gonzalez}, Guillermo},
        title = "{The stellar metallicity-giant planet connection}",
      journal = {\mnras},
         year = 1997,
        month = feb,
       volume = {285},
       number = {2},
        pages = {403-412},
          doi = {10.1093/mnras/285.2.403},
       adsurl = {https://ui.adsabs.harvard.edu/abs/1997MNRAS.285..403G}
}

@ARTICLE{Lau22a,
       author = {{Lau}, Mike Y.~M. and {Hirai}, Ryosuke and {Gonz{\'a}lez-Bol{\'\i}var}, Miguel and {Price}, Daniel J. and {De Marco}, Orsola and {Mandel}, Ilya},
        title = "{Common envelopes in massive stars: towards the role of radiation pressure and recombination energy in ejecting red supergiant envelopes}",
      journal = {\mnras},
         year = 2022,
        month = jun,
       volume = {512},
       number = {4},
        pages = {5462-5480},
          doi = {10.1093/mnras/stac049},
archivePrefix = {arXiv},
       eprint = {2111.00923},
 primaryClass = {astro-ph.SR},
       adsurl = {https://ui.adsabs.harvard.edu/abs/2022MNRAS.512.5462L}
}

@ARTICLE{Lau22b,
       author = {{Lau}, Mike Y.~M. and {Hirai}, Ryosuke and {Price}, Daniel J. and {Mandel}, Ilya},
        title = "{Common envelopes in massive stars II: The distinct roles of hydrogen and helium recombination}",
      journal = {\mnras},
         year = 2022,
        month = nov,
       volume = {516},
       number = {4},
        pages = {4669-4678},
          doi = {10.1093/mnras/stac2490},
archivePrefix = {arXiv},
       eprint = {2206.06411},
 primaryClass = {astro-ph.SR},
       adsurl = {https://ui.adsabs.harvard.edu/abs/2022MNRAS.516.4669L}
}

@ARTICLE{Ceillier17,
       author = {{Ceillier}, T. and {Tayar}, J. and {Mathur}, S. and {Salabert}, D. and {Garc{\'\i}a}, R.~A. and {Stello}, D. and {Pinsonneault}, M.~H. and {van Saders}, J. and {Beck}, P.~G. and {Bloemen}, S.},
        title = "{Surface rotation of Kepler red giant stars}",
      journal = {\aap},
         year = 2017,
        month = sep,
       volume = {605},
          eid = {A111},
        pages = {A111},
          doi = {10.1051/0004-6361/201629884},
archivePrefix = {arXiv},
       eprint = {1707.05989},
 primaryClass = {astro-ph.SR},
       adsurl = {https://ui.adsabs.harvard.edu/abs/2017A&A...605A.111C}
}

@ARTICLE{Rodrigez-Segovia25,
       author = {{Rodr{\'\i}guez-Segovia}, Nicol{\'a}s and {Ruiter}, Ashley J.},
        title = "{Population synthesis of hot subdwarf B stars with COMPAS: on the observed Galactic population}",
      journal = {\mnras},
         year = 2025,
        month = jun,
       volume = {539},
       number = {4},
        pages = {3273-3284},
          doi = {10.1093/mnras/staf710},
archivePrefix = {arXiv},
       eprint = {2505.05791},
 primaryClass = {astro-ph.SR},
       adsurl = {https://ui.adsabs.harvard.edu/abs/2025MNRAS.539.3273R}
}

@ARTICLE{Chabrier00,
       author = {{Chabrier}, Gilles and {Baraffe}, Isabelle},
        title = "{Theory of Low-Mass Stars and Substellar Objects}",
      journal = {\araa},
         year = 2000,
        month = jan,
       volume = {38},
        pages = {337-377},
          doi = {10.1146/annurev.astro.38.1.337},
archivePrefix = {arXiv},
       eprint = {astro-ph/0006383},
 primaryClass = {astro-ph},
       adsurl = {https://ui.adsabs.harvard.edu/abs/2000ARA&A..38..337C}
}

@ARTICLE{Zorotovic10,
       author = {{Zorotovic}, M. and {Schreiber}, M.~R. and {G{\"a}nsicke}, B.~T. and {Nebot G{\'o}mez-Mor{\'a}n}, A.},
        title = "{Post-common-envelope binaries from SDSS. IX: Constraining the common-envelope efficiency}",
      journal = {\aap},
         year = 2010,
        month = sep,
       volume = {520},
          eid = {A86},
        pages = {A86},
          doi = {10.1051/0004-6361/200913658},
archivePrefix = {arXiv},
       eprint = {1006.1621},
 primaryClass = {astro-ph.SR},
       adsurl = {https://ui.adsabs.harvard.edu/abs/2010A&A...520A..86Z}
}

@ARTICLE{De_Marco11,
       author = {{De Marco}, Orsola and {Passy}, Jean-Claude and {Moe}, Maxwell and {Herwig}, Falk and {Mac Low}, Mordecai-Mark and {Paxton}, Bill},
        title = "{On the {\ensuremath{\alpha}} formalism for the common envelope interaction}",
      journal = {\mnras},
         year = 2011,
        month = mar,
       volume = {411},
       number = {4},
        pages = {2277-2292},
          doi = {10.1111/j.1365-2966.2010.17891.x},
archivePrefix = {arXiv},
       eprint = {1010.4374},
 primaryClass = {astro-ph.SR},
       adsurl = {https://ui.adsabs.harvard.edu/abs/2011MNRAS.411.2277D}
}

@ARTICLE{Zorotovic22,
       author = {{Zorotovic}, Monica and {Schreiber}, MatthiasR},
        title = "{Close detached white dwarf + brown dwarf binaries: further evidence for low values of the common envelope efficiency}",
      journal = {\mnras},
         year = 2022,
        month = jul,
       volume = {513},
       number = {3},
        pages = {3587-3595},
          doi = {10.1093/mnras/stac1137},
archivePrefix = {arXiv},
       eprint = {2204.13715},
 primaryClass = {astro-ph.SR},
       adsurl = {https://ui.adsabs.harvard.edu/abs/2022MNRAS.513.3587Z}
}

@ARTICLE{Arancibia-Rojas24,
       author = {{Arancibia-Rojas}, Eduardo and {Zorotovic}, Monica and {Vu{\v{c}}kovi{\'c}}, Maja and {Bobrick}, Alexey and {Vos}, Joris and {Piraino-Cerda}, Franco},
        title = "{The mass range of hot subdwarf B stars from MESA simulations}",
      journal = {\mnras},
         year = 2024,
        month = feb,
       volume = {527},
       number = {4},
        pages = {11184-11197},
          doi = {10.1093/mnras/stad3891},
archivePrefix = {arXiv},
       eprint = {2312.09920},
 primaryClass = {astro-ph.SR},
       adsurl = {https://ui.adsabs.harvard.edu/abs/2024MNRAS.52711184A}
}
 % ##########################################################################################################

\begin{appendix}

\section{Evolution of the binding energy for the $1.0-2.0\Msun$ models}

	\begin{figure}[h]
		\centering
		\includegraphics[width=0.49\textwidth]{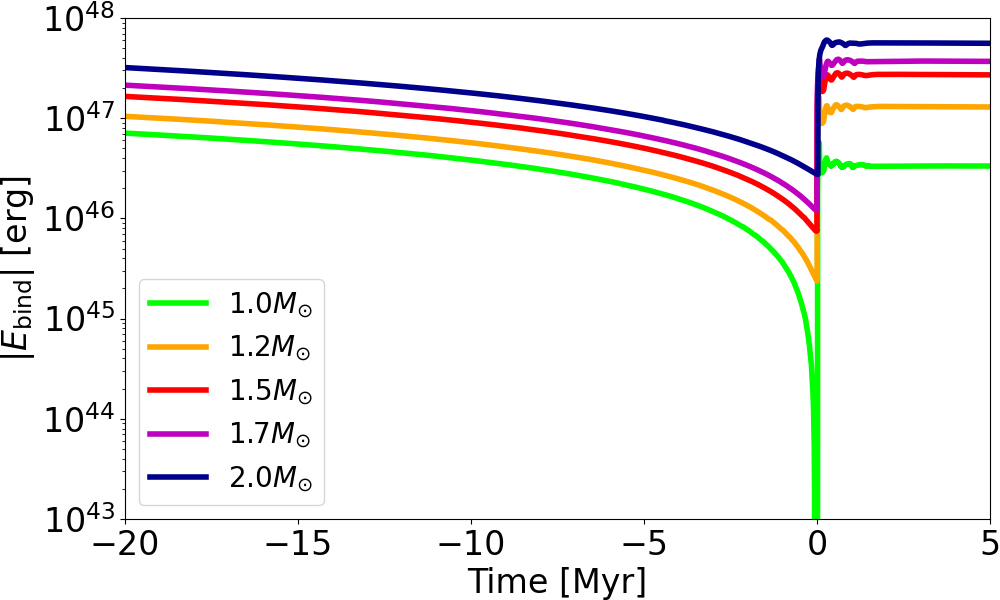}
				\caption{Evolution of $|E_\bind|$ for five stellar models with masses between 1.0 and $2.0\Msun$ from 20 Myr before to 5 Myr after the RGB tip. The $1.0\Msun$ falls outside the plotted range because its binding energy changes sign. All tracks have been synchronised on the RGB tip.}
		\label{fig:E bind multistar}
	\end{figure}

% #############################################################################################################

\section{Fit coefficients to the critical companion masses $M_p^\engl$ and $M_p^\eject$}
\label{sec:fate zone borders}

The companion mass $M_p^\engl$ that defines the boundary between engulfment and survival only depends on the initial separation $a$ and is independent of the energy-transfer efficiency $\alpha$.

\begin{table}[b]
    \centering
    \begin{tabular}{c|c|c|c|c|c}
    \hline
    $M_\star$ & $\alpha'$ & $\beta$ & $\gamma$ & $\delta$ & $\epsilon$ \\
    \hline
    1.2 & $2.05 10^{-4}$ & $4.16$ & $-1.43$ & $5.33$ & $-11.48$ \\
    1.5 & $1.83 10^{-3}$ & $3.68$ & $-1.17$ & $1.98$ & $-3.97$ \\
    1.7 & $1.94 10^{-3}$ & $3.799$ & $-1.153$ & $3.616$ & $-7.265$ \\
    2.0 & $7.74 10^{-3}$ & $3.757$ & $-0.795$ & $0.192$ & $-0.490$ \\
    \hline
    \end{tabular}
    \label{tab:fate_zones_engulf_coef}
    \caption{Fate zones fitting coefficient for $M_p^\engl$ (Eq.~\ref{eq:engulf limit}, in unit of Jupiter mass)}
\end{table}

\begin{table*}[b]
    \begin{tabular}{c|c|c|c|c|c|c|c}
    \hline
    $\alpha$ & $M_\star$ & $\alpha'$ & $\beta$ & $\gamma$ & $\delta$ & $\epsilon$ & $\zeta$ \\
    \hline
    1.0 & 1.2 & $-13.51$ & $104.0$ & -309.9 & 457.3 & -371.7 & 172.3 \\
    1.0 & 1.5 & $-4.56$ & $44.9$ & -169.4 & 317.1 & -330.3 & 191.2 \\
    1.0 & 1.7 & $-10.36$ & 90.0 & -306.7 & 520.8 & -480.5 & 242.4 \\
    1.0 & 2.0 & $-9.12$ & 87.6 & -326.0 & 597.4 & -580.9 & 296.9 \\
    \hline
    0.5 & 1.2 & 1.81 & -12.92 & 23.48 & 34.89 & -181.97 & 207.79 \\
    0.5 & 1.5 & 9.95 & -96.60 & 359.36 & -617.21 & 415.38 & 22.47 \\
    0.5 & 1.7 & -19.81 & 199.73 & -805.16 & 1639.44 & -1745.84 & 852.81 \\
    0.5 & 2.0 & 19.29 & -138.03 & 318.21 & -128.47 & -481.63 & 573.49 \\
    \hline
    0.2 & 1.2 & -12.15 & 112.38 & -411.02 & 763.85 & -809.92 & 497.03 \\
    0.2 & 1.5 & 1.66 & -15.49 & 60.44 & -110.92 & 6.53 & 217.87 \\
    0.2 & 1.7 & -88.38 & 1047.53 & -4925.46 & 11493.37 & -13389.78 & 6366.01 \\
    0.2 & 2.0 & -2.42 & 22.38 & -70.18 & 123.09 & -295.77 & 500.00 \\
    \hline
    \label{tab:fate_zones_eject_coef}
    \end{tabular}
    \caption{Fitting coefficient for $M_p^\eject$ (Eq.~\ref{eq:eject limit}, in unit of Jupiter mass)}
\end{table*}

\section{Evolution of the ratio $E_\orb / |E_\bind|$ for the 1.2 and $2.0\Msun$ models}

    \begin{figure}[h]
		\centering
		\includegraphics[width=0.46\textwidth]{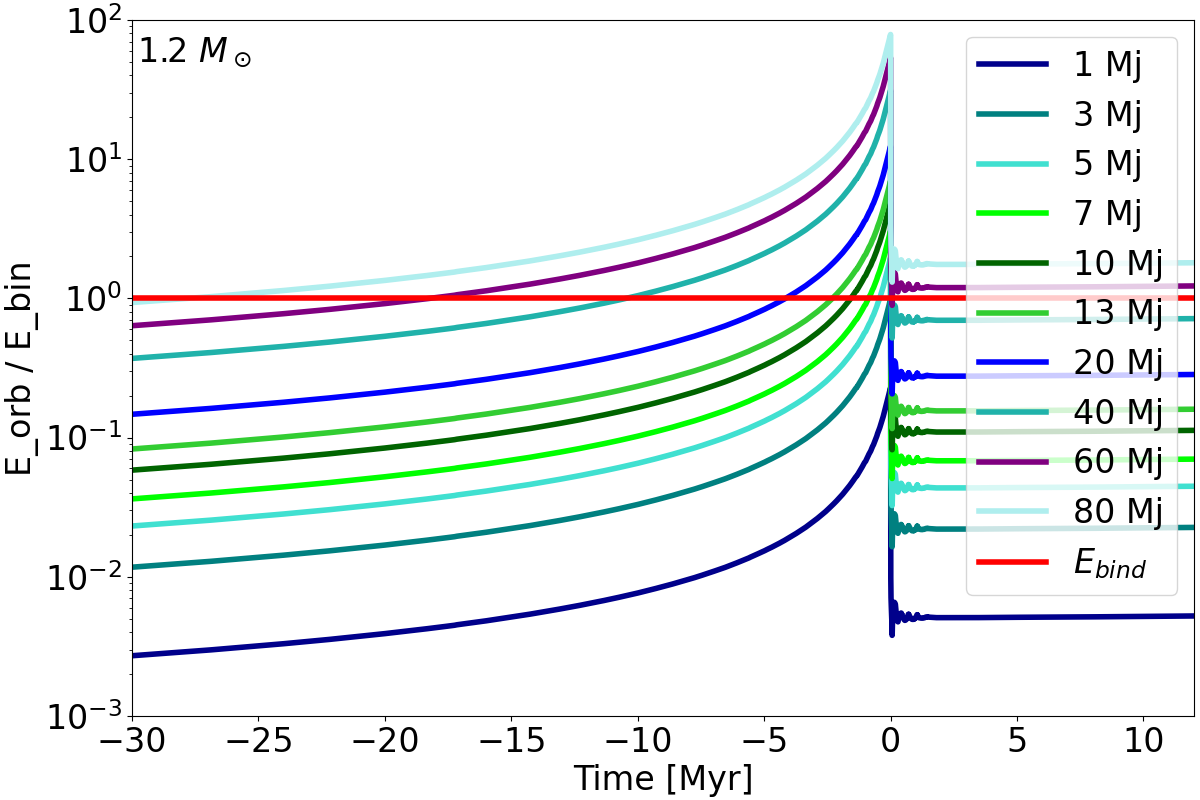}
		\includegraphics[width=0.46\textwidth]{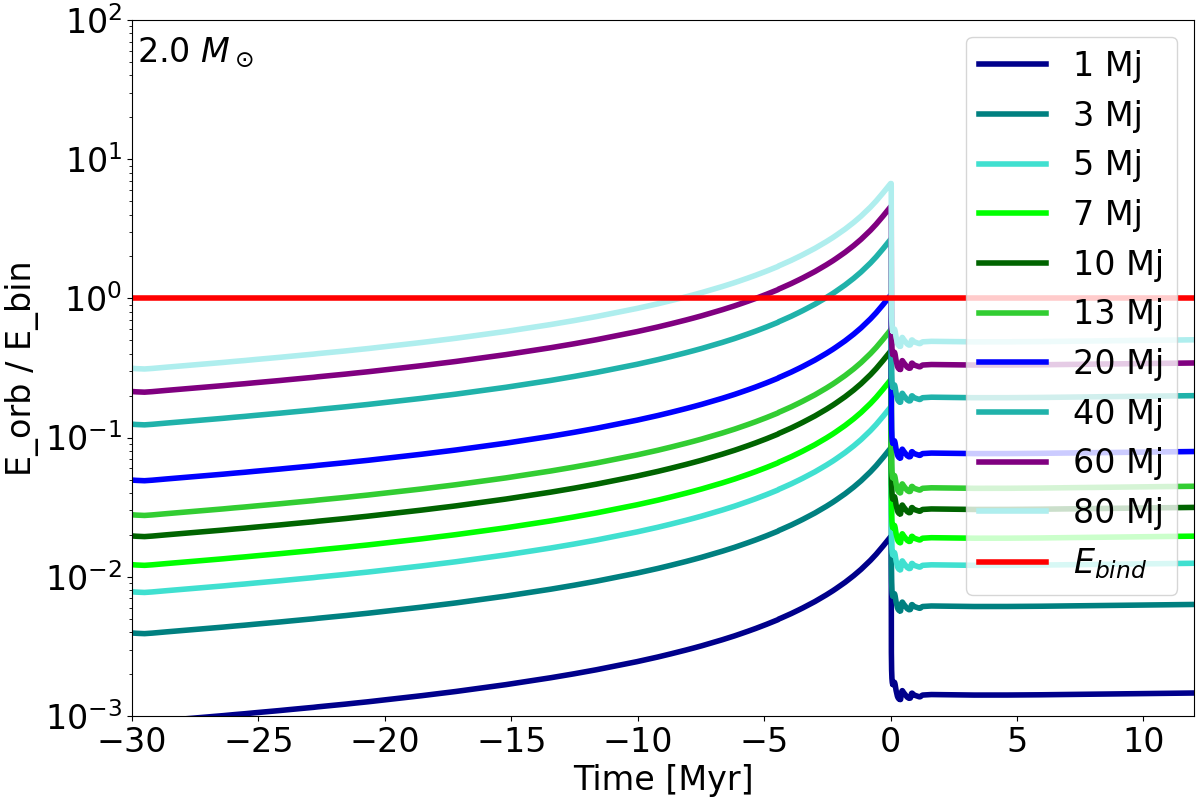}
		\caption{Orbital energy released by disrupted bodies of masses between 1 and $80\,\Mj$ (all with a radius of $1\,\Rj$) when engulfed by stars of mass 1.2 and $2.0\Msun$, relative to the binding energy of the envelope $E_\bind$. The origin of time is defined at the RGB-tip.}
		\label{fig: E balance allstar}
	\end{figure}

\section{Additional results on envelope ejection and core mass at ejection}
\label{apd:occurrences}

\begin{figure*}[h!]
        \begin{tabular}{|c|c|c|}
        \includegraphics[width=0.31\textwidth]{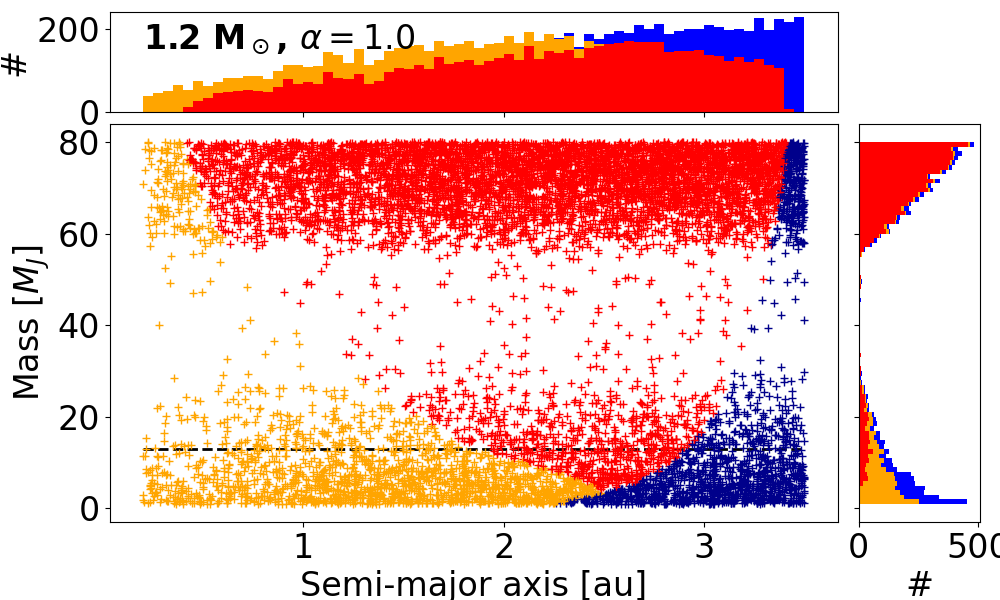} & \includegraphics[width=0.31\textwidth]{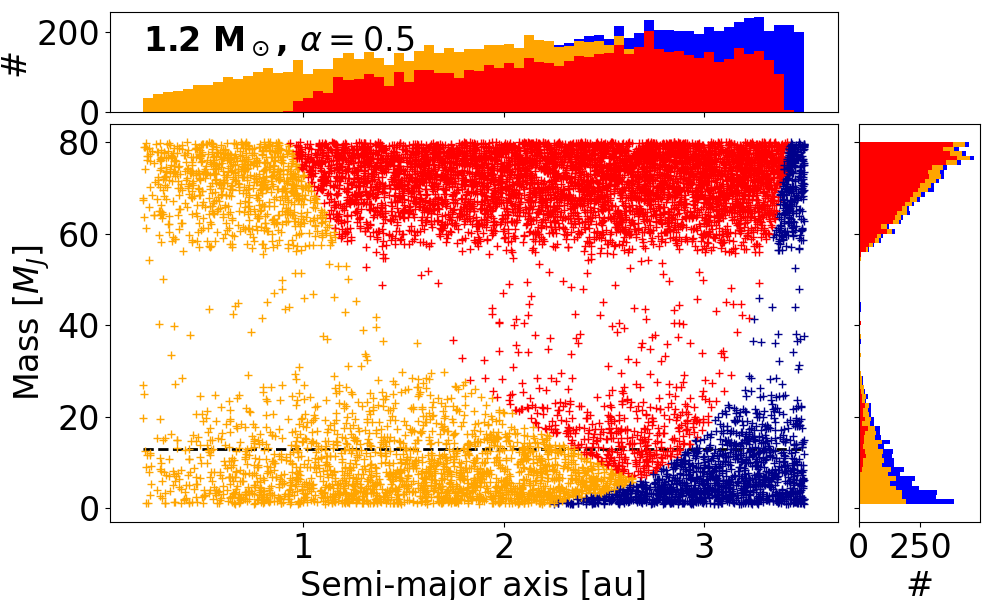} & \includegraphics[width=0.31\textwidth]{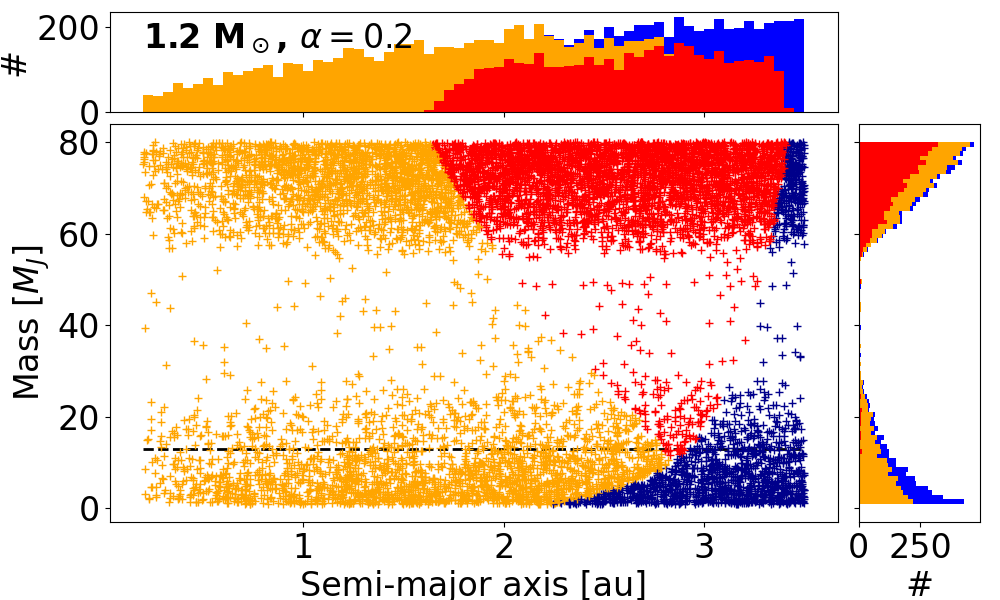} \\
        
        \includegraphics[width=0.31\textwidth]{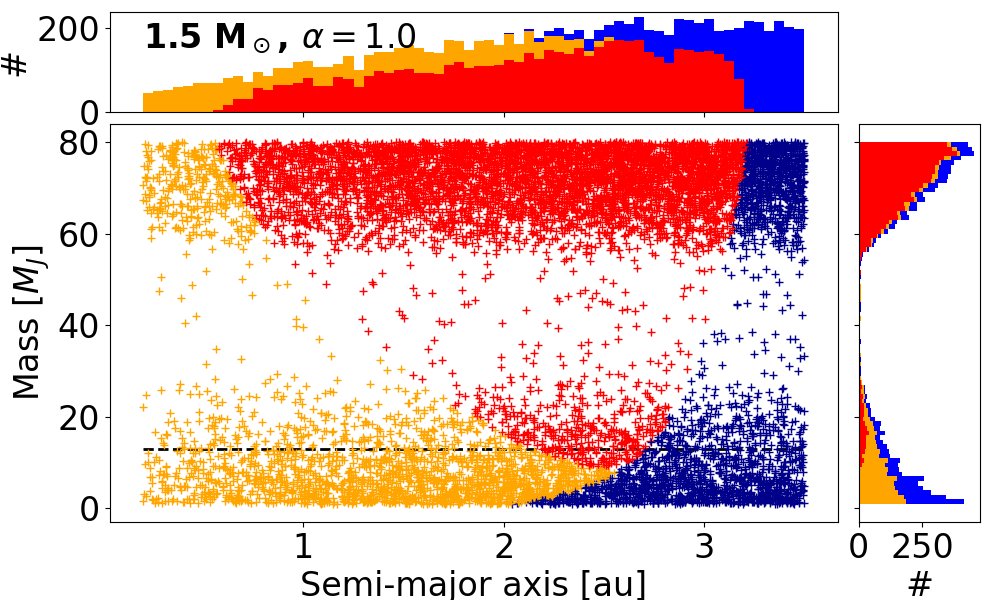} & \includegraphics[width=0.31\textwidth]{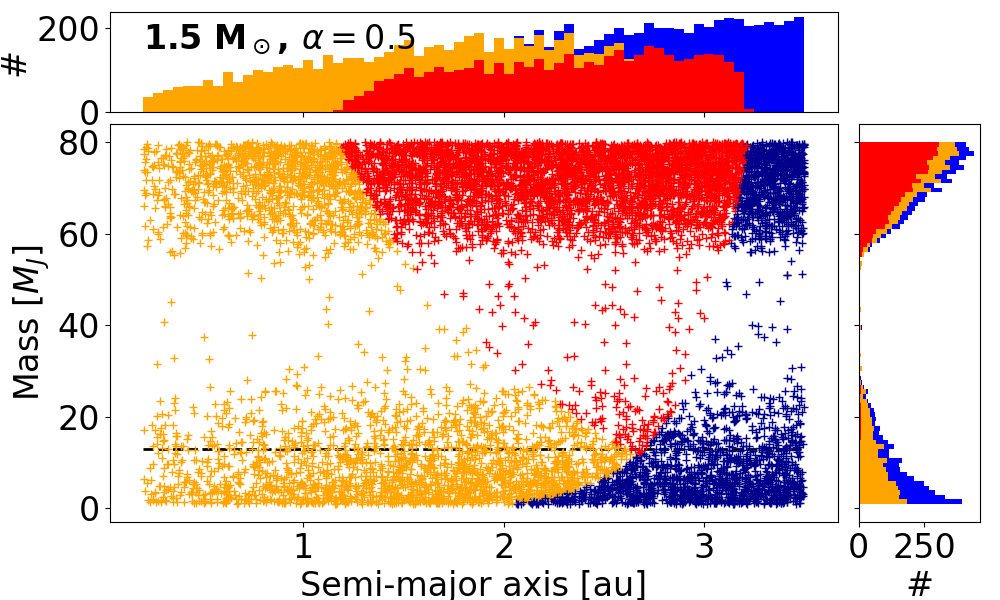} & \includegraphics[width=0.31\textwidth]{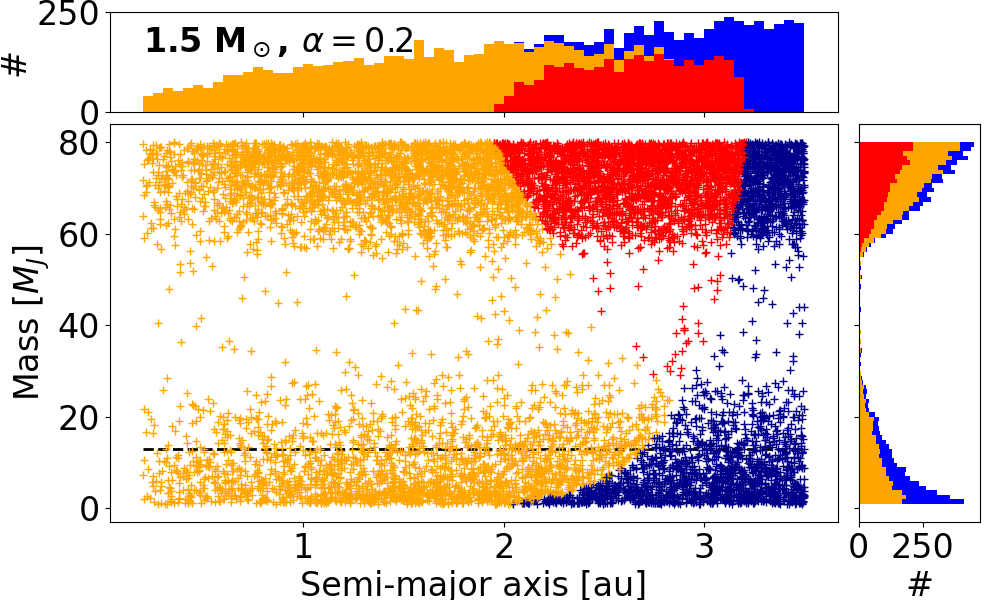} \\
        
        \includegraphics[width=0.31\textwidth]{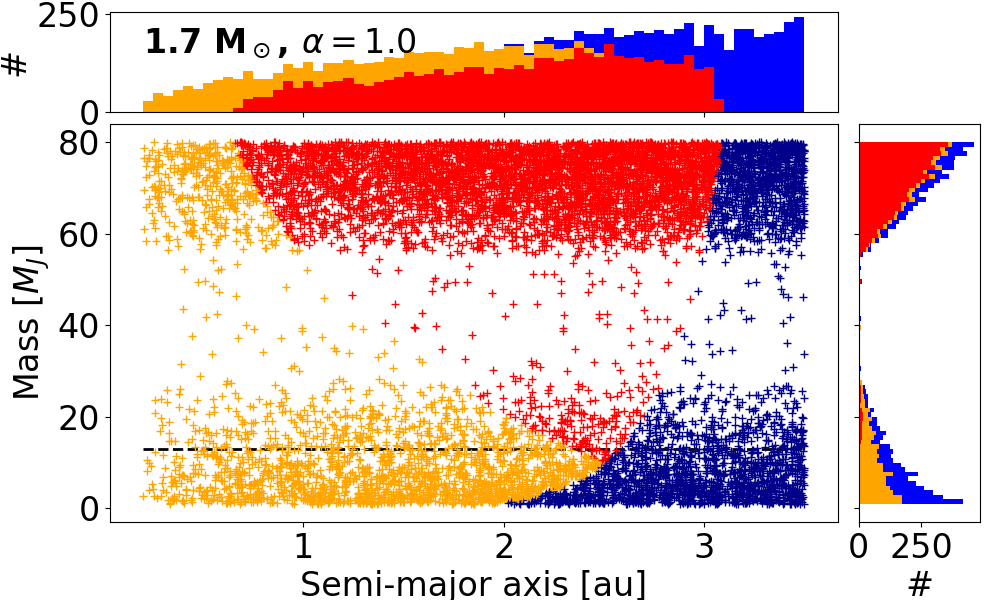} & \includegraphics[width=0.31\textwidth]{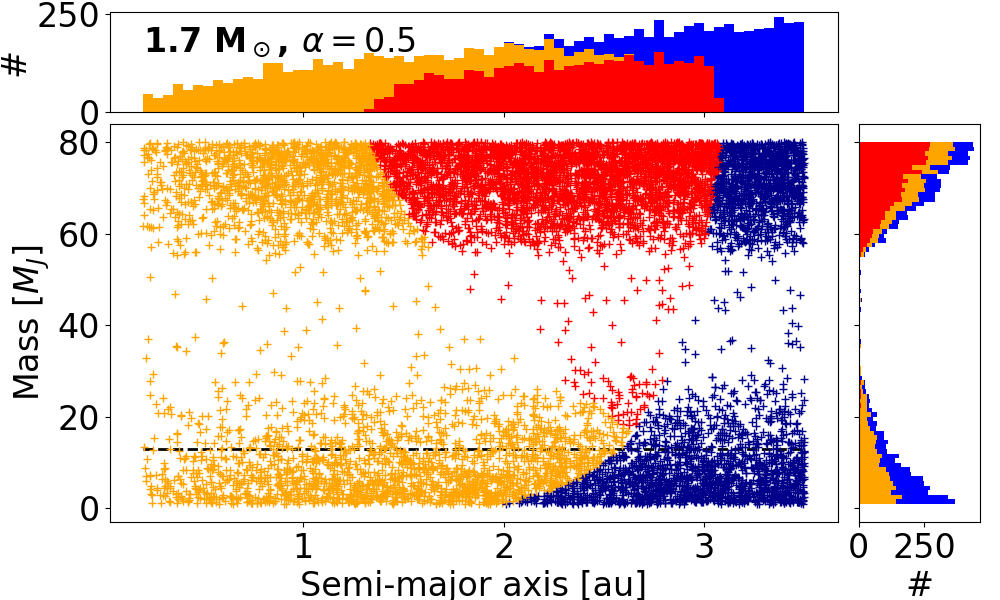} & \includegraphics[width=0.31\textwidth]{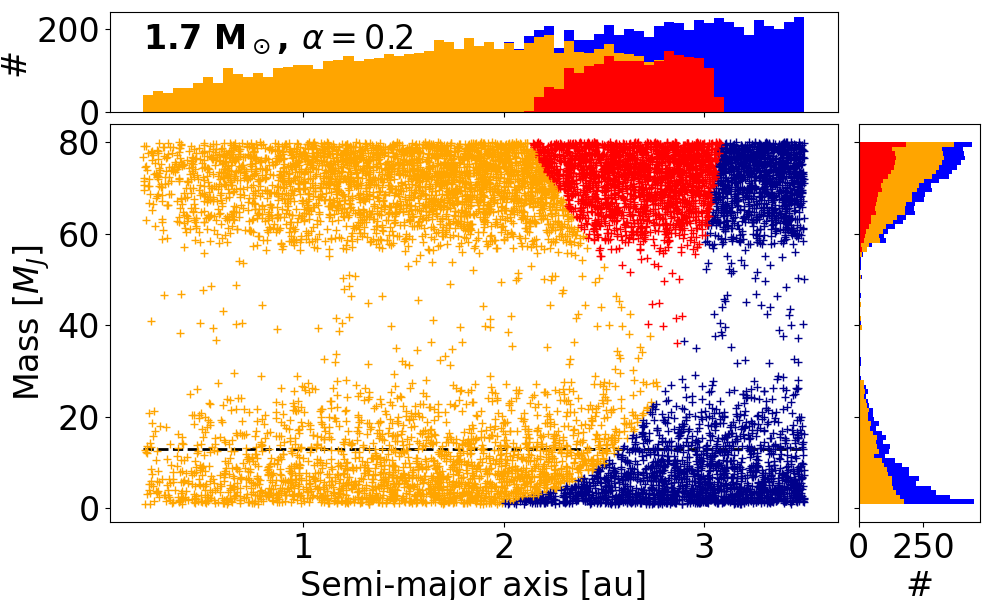} \\
        
        \includegraphics[width=0.31\textwidth]{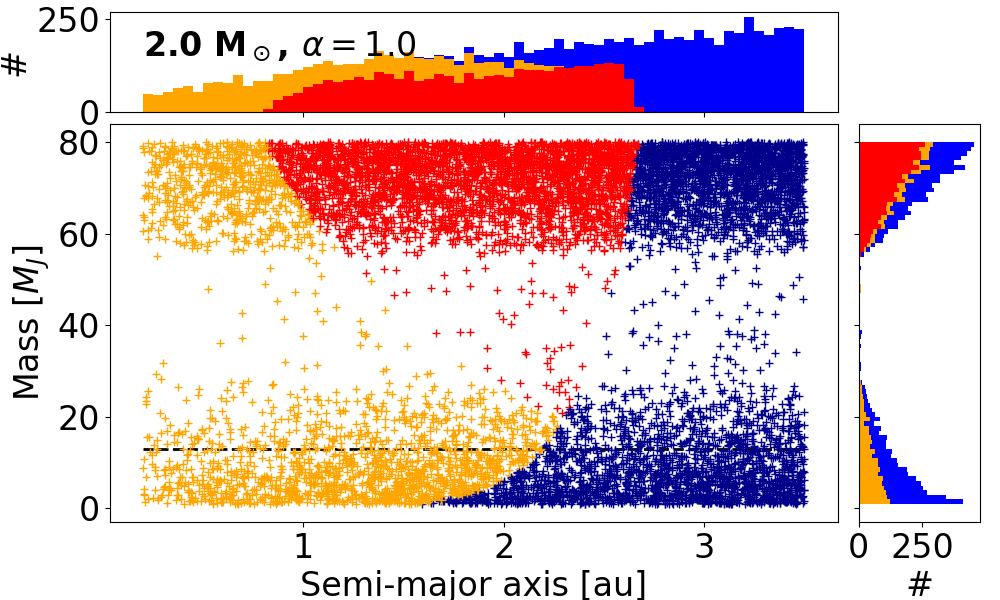} & \includegraphics[width=0.31\textwidth]{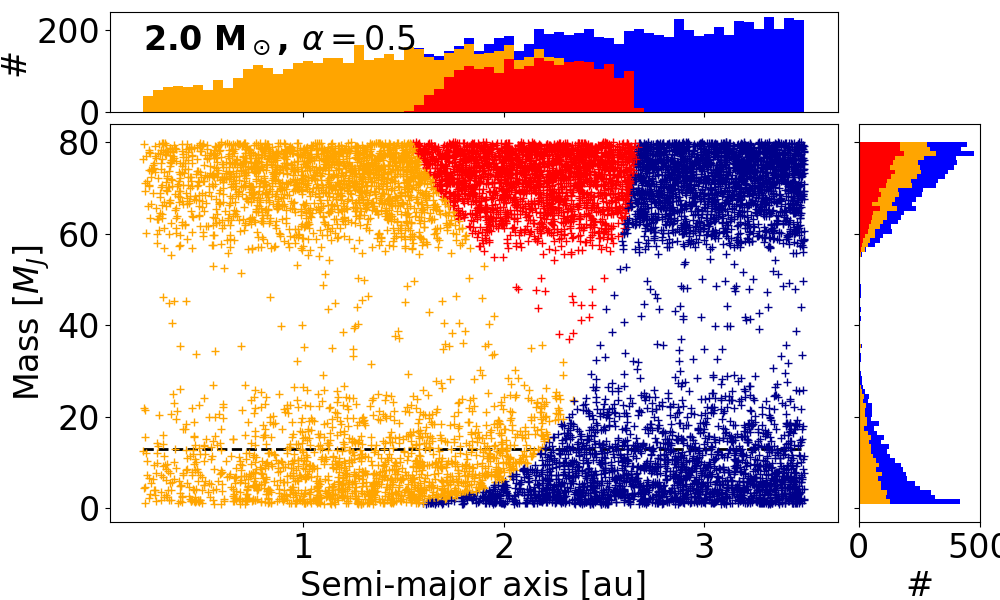} & \includegraphics[width=0.31\textwidth]{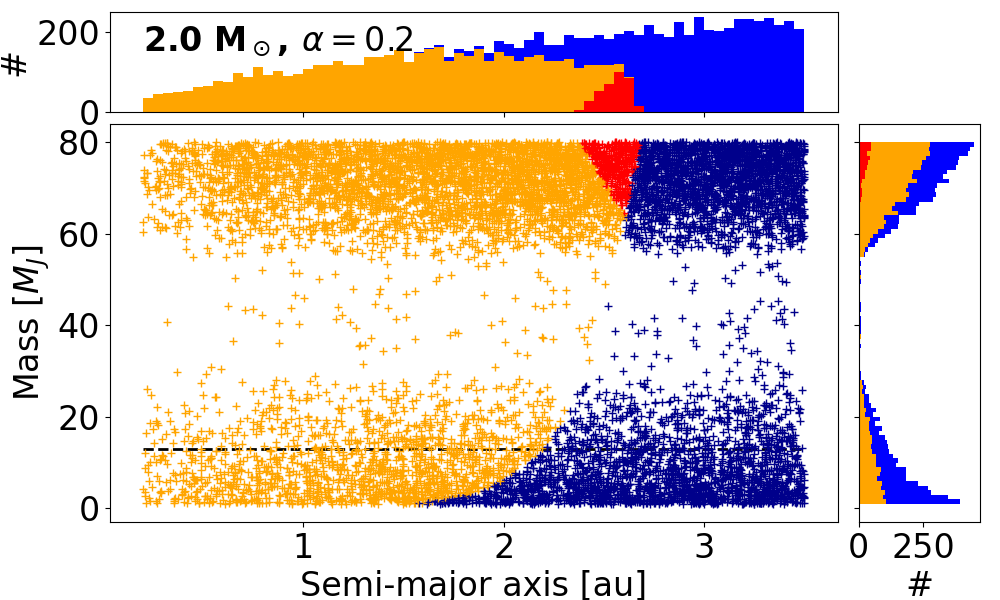} \\
        \end{tabular}
        \caption{Same as Fig.~\ref{fig:generated Ms1.5} but for stellar masses of 1.2 to $2.0\Msun$ and energy transfer efficiencies of 0.2, 0.5 and 1.}
		\label{fig:generated allstar}
	\end{figure*}

    \begin{figure}[h]
		\centering
		\includegraphics[width=0.46\textwidth]{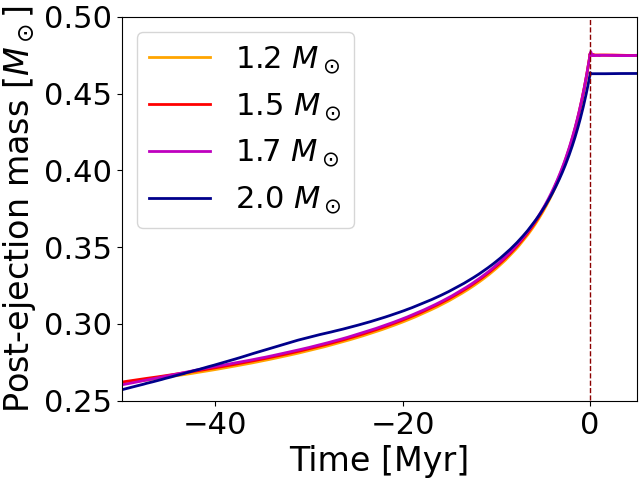}
		\caption{Core mass reached by the STAREVOL models at the time of envelope ejection, for ejection times spanning the final 50 Myr before the RGB tip. The origin of time is defined at the RGB tip and highlighted with a dashed vertical line.}
		\label{fig:Post-ejection_mass}
	\end{figure}

\end{appendix}

\end{document}